\documentclass[referee,a4paper,12pt,traditabstract]{swsc} 

\usepackage{graphicx}
\usepackage{txfonts}
\usepackage{subfigure}
\usepackage{epstopdf}
\usepackage{pdfpages}
\usepackage[mathlines]{lineno}
\usepackage[authoryear,round]{natbib}
\usepackage[backref]{hyperref}
\usepackage{url}
\usepackage[utf8]{inputenc}
\usepackage{lscape}
\usepackage{enumitem}
\usepackage[utf8]{inputenc}
\usepackage[greek.polutoniko,english]{babel}
\hypersetup{colorlinks=true,citecolor=cyan,urlcolor=cyan,linkcolor=blue}

\begin{document}

%\begin{linenumbers}  

   \title{The Flare Likelihood and Region Eruption Forecasting (FLARECAST) Project: Flare forecasting in the big data \& machine learning era}
   
   \titlerunning{The EU FLARECAST Project}

   \authorrunning{Georgoulis, Bloomfield, Piana \& the FLARECAST team}

   % references for institutes allow rearranging the institutes'
   % numbering easily if required by changes in the authors list (EB)
   \author{M. K. Georgoulis\inst{\ref{i:aa}},
           D. S. Bloomfield\inst{\ref{i:unn},\ref{i:tcd}},
           M. Piana\inst{\ref{i:unige},\ref{i:cnr}}, 
           A. M. Massone\inst{\ref{i:unige},\ref{i:cnr}},
           M. Soldati\inst{\ref{i:fhnw}},
           P. T. Gallagher\inst{\ref{i:dias},\ref{i:tcd}},
           E. Pariat\inst{\ref{i:cnrs}},
           N. Vilmer\inst{\ref{i:cnrs}},
           E. Buchlin\inst{\ref{i:psud}},
           F. Baudin\inst{\ref{i:psud}},
           A. Csillaghy\inst{\ref{i:fhnw}},
           H. Sathiapal\inst{\ref{i:fhnw}},
           D. R. Jackson\inst{\ref{i:metoffice}},
           P. Alingery\inst{\ref{i:psud}},
           F. Benvenuto\inst{\ref{i:unige}},
           C. Campi\inst{\ref{i:cnr},\ref{i:unipd}},
           K. Florios\inst{\ref{i:aa},\ref{i:teia}},
           C. Gontikakis\inst{\ref{i:aa}},
           C. Guennou\inst{\ref{i:cnrs}},
           J. A. Guerra\inst{\ref{i:tcd},\ref{i:villanova}},
           I. Kontogiannis\inst{\ref{i:aa},\ref{i:aip}},
           V. Latorre\inst{\ref{i:unige}},
           S. A. Murray\inst{\ref{i:tcd},\ref{i:dias}},
           S.-H. Park\inst{\ref{i:tcd},\ref{i:nagoya}},
           S. von Stachelski\inst{\ref{i:fhnw}},
           A. Torbica\inst{\ref{i:fhnw}},
           D. Vischi\inst{\ref{i:fhnw}},
           \and 
           M. Worsfold\inst{\ref{i:metoffice}}
          }

   \institute{RCAAM of the Academy of Athens, 11527 Athens, Greece\label{i:aa}\\\email{\href{mailto:manolis.georgoulis@academyofathens.gr}{manolis.georgoulis@academyofathens.gr}}\thanks{corresponding author}
 %             \and
 %             Department of Physics \& Astronomy, Georgia State University, Atlanta, GA 30303, USA\label{i:georgia}\\\email{\href{mailto:manolis.phy-astr.gsu.edu}{manolis.georgoulis@phy-astr.gsu.edu}}
              \and
              Department of Mathematics, Physics \& Electrical Engineering, Northumbria University, Newcastle upon Tyne, NE1 8ST, UK\label{i:unn}
              \and
              School of Physics, Trinity College Dublin, Dublin 2, Ireland\label{i:tcd}
              \and 
              Dipartimento di Matematica, Universit\`a di Genova, via Dodecaneso 35 16146 Genova, Italy\label{i:unige}
              \and
              CNR - SPIN Genova, via Dodecaneso 33 16146 Genova, Italy\label{i:cnr}
              \and
              University of Applied Sciences \& Arts Northwestern Switzerland, 5210, Windisch, Switzerland\label{i:fhnw}
              \and 
              School of Cosmic Physics, Dublin Institute for Advanced Studies, Dublin, D02 XF85, Ireland\label{i:dias}
              \and
              LESIA, Observatoire de Paris, Universit\'e PSL, CNRS, Sorbonne Universit\'e, Universit\'e de Paris, France\label{i:cnrs}
              \and 
              Université Paris-Saclay, CNRS,  Institut d'Astrophysique Spatiale, 91405, Orsay, France\label{i:psud} % [EB has been changed to reflect new name]
              \and
              Met Office, Exeter, UK\label{i:metoffice}
              \and
              Dipartimento di Matematica ``Tullio Levi-Civita'', Universit\`a di Padova, Padova, Italy\label{i:unipd}
              \and
              Department of Chemical Engineering, National Technical University of Athens, Athens, Greece\label{i:teia}
              \and
              Physics Department, Villanova University, 800 E Lancaster Ave. Villanova, PA 19085, USA.\label{i:villanova}
              \and
              Leibniz-Institut f\"ur Astrophysik Potsdam (AIP), Potsdam, Germany\label{i:aip}
              \and 
              Institute for Space-Earth Environmental Research, Nagoya University, Nagoya, Japan\label{i:nagoya}
              }

%%   \date{Received September 15, 1996; accepted March 16, 1997}

  % \abstract{}{}{}{}{}        %% uncomment if structured abstract is desired
 %% 5 {} token are mandatory
 
  \abstract{
  The European Union funded the FLARECAST project, that ran from January 2015 until February 2018. 
  FLARECAST had a research-to-operations (R2O) focus, and accordingly introduced several innovations into the discipline of solar flare forecasting. 
  FLARECAST innovations were: first, the treatment of hundreds of physical properties viewed as promising flare predictors on equal footing, extending multiple previous works; second, the use of fourteen (14) different machine learning techniques, also on equal footing, to optimize the immense Big Data parameter space created by these many predictors; third, the establishment of a robust, three-pronged communication effort oriented toward policy makers, space-weather stakeholders and the wider public. FLARECAST pledged to make all its data, codes and infrastructure openly available worldwide. 
  The combined use of 170+ properties (a total of 209 predictors are now available) in multiple machine-learning algorithms, some of which were designed exclusively for the project, gave rise to changing sets of best-performing predictors for the forecasting of different flaring levels, at least for major flares. 
  At the same time, FLARECAST reaffirmed the importance of rigorous training and testing practices to avoid overly optimistic pre-operational prediction performance. 
  In addition, the project has (a) tested new and revisited physically intuitive flare predictors and (b) provided meaningful clues toward the transition from flares to eruptive flares, namely, events associated with coronal mass ejections (CMEs). 
  These leads, along with the FLARECAST data, algorithms and infrastructure, could help facilitate integrated space-weather forecasting efforts that take steps to avoid effort duplication. 
  In spite of being one of the most intensive and systematic flare forecasting efforts to-date, FLARECAST has not managed to convincingly lift the barrier of stochasticity in solar flare occurrence and forecasting: solar flare prediction thus remains inherently probabilistic.
  }

   \maketitle
%%
%%________________________________________________________________

\begin{tabular}{l r r r r r r r r r l }
\textgreek{\small{<O <'hlios o>u m'onon... n'eos >ef' <hm'erh|}} 
&  & & & & & & & &  & \texttt{\small{The Sun is young every day,}}\\
\textgreek{\small{>est'in, >all' >ae'i n'eos suneq\~ws.}} 
&  & & & & & & & &  & \texttt{\small{incessantly and eternally.}}\\
&  & & & & & & & &  & \\
\textgreek{\small{>Hr'akleitos, $\sim$500 P.Κ.Ε.}} & & & & & & & & & &  
\texttt{\small{Heraclitus, $\sim$500 BCE}}\\
\end{tabular}

\section{Introduction}
\label{sec:intro}
The first decades of the 21st century have seen the transformative effect of the ever-increasing, widespread use of wireless technologies. 
Enhanced by the equally expansive use of the internet, these technologies have claimed, and are expected to continue claiming, a crucial part of our everyday routine, from services to communications and from information to edutainment. 
Space-based satellite technologies have also been instrumental in establishing wireless capabilities to a degree that few could predict, even by the standards of the late 20th century. 

When relying on space, however, it is imperative to keep in mind the decisive effects of our magnetically active star, the Sun. 
Simultaneously with the expansion of human capabilities came increased awareness of the adverse effects of space weather, namely, the short-term (hours to days) impact of solar magnetic activity, from the fast solar wind spewed by extended coronal holes to the storm-like transport of solar eruptions through the entire heliosphere. 
Electromagnetic and particulate emission from solar eruptions can cause anything from short-lived, relatively unimpactful disruptions to major damage in satellite infrastructure, on top of the biological hazards they pose to exposed humans in space conditions, either during extravehicular activities or in future manned space travel and missions to Moon and Mars (see, for example, ESA's\footnote{All abbreviations and acronyms used hereafter are explained in Appendix \ref{app4}.} Moon Village and NASA's Artemis  Programs).

\begin{figure}
    \centering
    \includegraphics[width=0.80\textwidth]{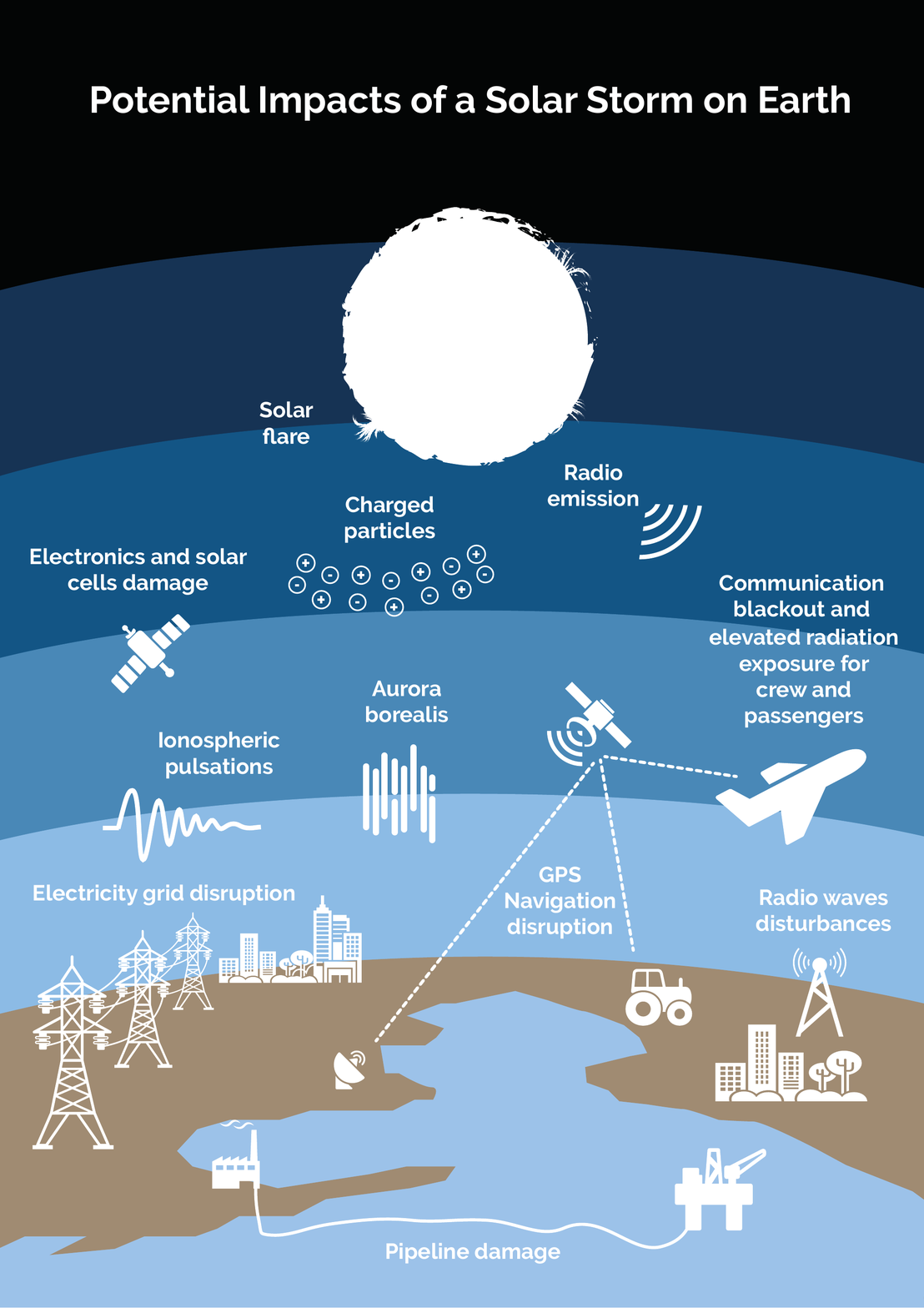}
    \caption{A graphical representation of extreme space weather and its effects on contemporary technology and infrastructure. Credit: FHNW.}
    \label{fig:fhnw}
\end{figure}

Exceptional solar flares and eruptive manifestations, up to the first flare observed by \citet{car59}, are among deep-space phenomena whose repercussions go past the ionosphere, reaching down to aviation altitudes and even to Earth's surface. 
Figure~\ref{fig:fhnw} portrays this impact in an image produced by the University of Applied Sciences and Arts of Northwestern Switzerland (FHNW) partner of the FLARECAST Consortium. 
There, one sees ramifications spanning from what we already knew before the space age (i.e., the aurora, long-range electrical power networks or radar disruptions) to any applications that GPS or Galileo enable. 
During the space age we have seen some solar eruptions that have caused major effects in May 1967, March 1989, and October-November 2003, although solar flares associated with these eruptions were arguably smaller than the Carrington flare \citep{cliver_dietrich13}. 
However, while cruising on the far-side of the Sun in July 2012, the STEREO-A spacecraft detected the transit of an extreme solar eruption that was characterized as a Carrington-scale event \citep{baker_etal13}. 
Geospace was spared from that eruption but, statistically, in future events it will not be \citep[see, e.g.,][]{riley_love17}.

The staggering short- and long-term financial impact of extreme space weather has been delineated in a series of recent works \citep[e.g.,][]{macalester2014,oughton_etal17,eastwood_etal18}, as well as in governmental guidelines and action plans, such as the \emph{US National Space Weather Strategy and Action Plan} (2015, 2019{\footnote{Available at \url{https://www.whitehouse.gov/wp-content/uploads/2019/03/National-Space-Weather-Strategy-and-Action-Plan-2019.pdf}}}) and the \emph{UK Space Weather Preparedeness Strategy} (2015{\footnote{Available at \url{https://www.gov.uk/government/publications/space-weather-preparedness-strategy}}}). 
Of particular interest, however, is the 2008 National Research Council's \emph{Severe Space Weather Events: Understanding Societal and Economic Impacts Workshop Report} \citep{NAS08} that portrays in its Figure 3.1 the nonlinear inner workings and interconnections of sectors comprising our societal fabric. 
Like domino bricks, if any of these sectors fails due to extreme space weather, the ramifications will be hard to imagine and even harder to mitigate. 
The reliable forecasting of extreme space weather, therefore, upgrades to a major challenge of our times. 

Energetic events accompanying major solar eruptions are by far the main agents of extreme space weather. These events comprise three distinct aspects: solar flares, CMEs, and SEP events. 
A reliable forecasting, therefore, should encompass three very different forecast efforts with unique characteristics and challenges. In solar flare prediction, that is the topic of discussion in this work, there is no early warning for the flares' X- and $\gamma$-ray photons. Only a slim window of $\sim10-12$~min exists for flare-accelerated SEPs, if any \citep{haggerty_roelof02,rust_etal08}. 
To address the lack of advance warning, therefore, major solar flares --and flare-related, impulsive SEP events, by extension-- need to be predicted well before their occurrence (i.e., several hours to $1-2$~days in advance). 
There are currently significant shortcomings in our flare forecasting ability, as sections below will show.
In addition, CMEs, particularly the fastest ones that are stemming from solar ARs and are virtually associated one-to-one to major flares \citep[][but see \cite{liu_etal16} for a exceptional active region]{yashiro_etal05}, should ideally be predicted along with flares and SEPs \citep[e.g.,][]{anastasiadis_etal17}. 
There is a window of inner-heliospheric transit ranging between $\sim20$~hr and $2-3$~days after the initial, near-Sun detection of CMEs until they reach geospace. 
If Earth-directed, their arrival time and geoeffectiveness (i.e., their potential ability to trigger a geomagnetic storm) should also be predicted in advance \citep[e.g.,][]{mostl_etal14,mays_etal15}. 
Finally, CME-shock-accelerated SEP events may arrive at geospace hours after the source solar eruption or, in the worst-case scenario, even before the CME registers in near-Sun height-time diagrams \citep[][and references therein]{reames_17,malandraki_etal18}. 
We also need to know in advance the temporal profile of the SEP event and its peak flux or fluence, particularly for proton energy channels exceeding $50-100$~MeV, as per NOAA guidelines and recently defined benchmarks\footnote{See the US Space Weather Phase 1 Benchmarks at \url{https://www.whitehouse.gov/wp-content/uploads/2018/06/Space-Weather-Phase-1-Benchmarks-Report.pdf}}. 

Although an ultimate goal, we still seem to be far from achieving an integrated platform for the prediction of all extreme space weather manifestations. 
Among them, solar flare prediction has historically been humanity's first stride. 
Since the 1980s, there have been persistent efforts toward flare prediction introducing a wealth of physical, semi-empirical or empirical AR properties and proxies that have been claimed to hold a flare-predictive capability. 
A short, non-exhaustive review of these properties, including an effort to group them into different categories, appears in \citet{georgoulis12}.

However, earlier efforts \citep[e.g.,][]{leka_barnes03,leka_barnes07,barnes_leka08} 
aiming to assess the relative performance of these properties indicated that, on one hand, none was solely capable to predict flares reliably while, on the other hand, 
when a capability to simultaneously test multiple properties was achieved, the predictive information contained in the full property set was highly redundant. 
%\mg{This became further evident when machine learning methods were put to work for flare prediction} \citep[][and several works that followed]{qahwaji_etal07,bobra_couvidat15}.
The first comparative evaluation of prominent flare-predictive properties and methodologies, undertaken by \citet{barnes_etal16}, established that no single method clearly outperformed the others. 
This and other initial findings (e.g,, the predictive value of timeseries, previous flare history) were further solidified by collaborative work on operational forecasts by \citet{leka_etal19a,leka_etal19b} and \citet{park_etal20}.

Meanwhile, the explosive increase in computing power spearheaded critical advances in computer science, in an already existing Big Data ecosystem facilitated by the wealth of ground- and space-based solar observations since the mid-1990s. 
Data mining and the advent of machine learning
%artificial intelligence, manifested initially by machine-learning methodologies, 
eventually led to the first application of a SVM and neural networks in flare forecasting 
\citep{qahwaji_etal07}. Several seminal works followed thereafter \citep{qahwaji_etal08,li_etal08,song_etal09,yu_etal09,bobra_couvidat15,muranushi_etal15,nishizuka_etal17} and the list is ever-expanding.
Today, we know that solar flare forecasting -- and space weather forecasting, in general -- should be viewed as an interdisciplinary effort, with a potentially critical contribution from machine learning \citep[][and references therein]{camporeale_etal18}, albeit not without open challenges impeding progress \citep{camporeale19}.

In this continuously and rapidly evolving landscape, a Consortium of nine institutes spanning over six European countries took advantage of the EU Horizon-2020 2014 PROTEC-2014 opportunity to propose the FLARECAST project. 
Having all the above in mind, FLARECAST pledged {\it to develop an advanced solar flare prediction system based on automatically extracted physical properties of solar ARs, coupled with state-of-the-art machine learning solar flare prediction methods and validated using the most appropriate forecast verification measures}. 
FLARECAST featured three top-level objectives, namely, one scientific, one devoted to the R2O philosophy and one engaging in communication. 
In particular, FLARECAST proposed:
\begin{itemize}
    \item In terms of science, to understand the drivers of solar flare activity and improve flare forecasting.
    \item In terms of R2O, to provide a globally accessible flare forecast service that facilitates expansion. 
    \item In terms of communication, to engage with space weather end users, inform stakeholders and policy makers, and educate the broader public on solar flares and space weather in general.
\end{itemize}
%The FLARECAST Consortium consisted of interdisciplinary teams based in   
%\begin{itemize}
%    \item Academy of Athens, Greece;
%    \item University of Northumbria at Newcastle, UK; 
%    \item Trinity College of the University of Dublin, Ireland; 
%    \item Universita di Genova, Italy; 
%    \item Consiglio Nazionale delle Ricerche, Italy; 
%    \item Centre National de la Recerche Scientifique, France; 
%    \item Institut d' Astrophysique Spatiale of the Universite Paris-Sud, France; 
%    \item Facchochschule Nordwestschweiz, Switzerland; 
%    \item Met Office, UK. 
%\end{itemize}

This collective work summarizes FLARECAST's most significant findings and conclusions in all three objectives above, along with the peer-reviewed publications that occurred in its course. 
Section~\ref{s2} describes the methodology followed throughout the project, while Section~\ref{s3} elaborates on the tasks of data handling and monitoring. 
Section~\ref{s4} discusses the FLARECAST performance verification strategy, while Section~\ref{sec:WP6} briefly describes the project's scientific results and future exploration component. 
Section~\ref{s6} encapsulates the main conclusions of the project's three top-level objectives (i.e., science, operations, communication) along with lessons learned in the course of the project. 
Finally, in a series of Appendices we provide detailed instructions on accessing FLARECAST data, codes and infrastructure (Appendix~\ref{app1}), key results from the FLARECAST Users Survey (Appendix~\ref{app2}; see Section~\ref{sec:dissem} for a relevant discussion), the list of refereed publications related to and acknowledging the FLARECAST project (Appendix~\ref{app3}) and, finally, a list of acronyms and abbreviations used in this paper (Appendix~\ref{app4}). 

All things considered, as commented by an attendee of one of the FLARECAST Stakeholders Workshops, \emph{``the real fun starts now''}; we expect a number of future works that will take advantage of and exploit the FLARECAST products. 
These may not be restricted to flare prediction, as (i) the volume of metadata provided by the processing of the 
NRT SHARP data product \citep{bobra_etal14} is substantial and (ii) as per EU's OpenAIRE initiative (\url{https://www.openaire.eu}), all data, codes  and infrastructure of the project are openly available to any interested individual or team worldwide. 
Ideally, then, one might view the FLARECAST infrastructure as a vehicle for a future integrated space weather prediction platform. 
Comprehensive and diverse material on the FLARECAST project can be found in the openly accessible FLARECAST website \url{http://flarecast.eu} (see also Fig.~\ref{fig:website} in Section~\ref{s2:comm} for a top-level structure of this website).

\section{The FLARECAST Approach}
\label{s2}
\begin{figure}
    \centering
    \includegraphics[width=\textwidth]{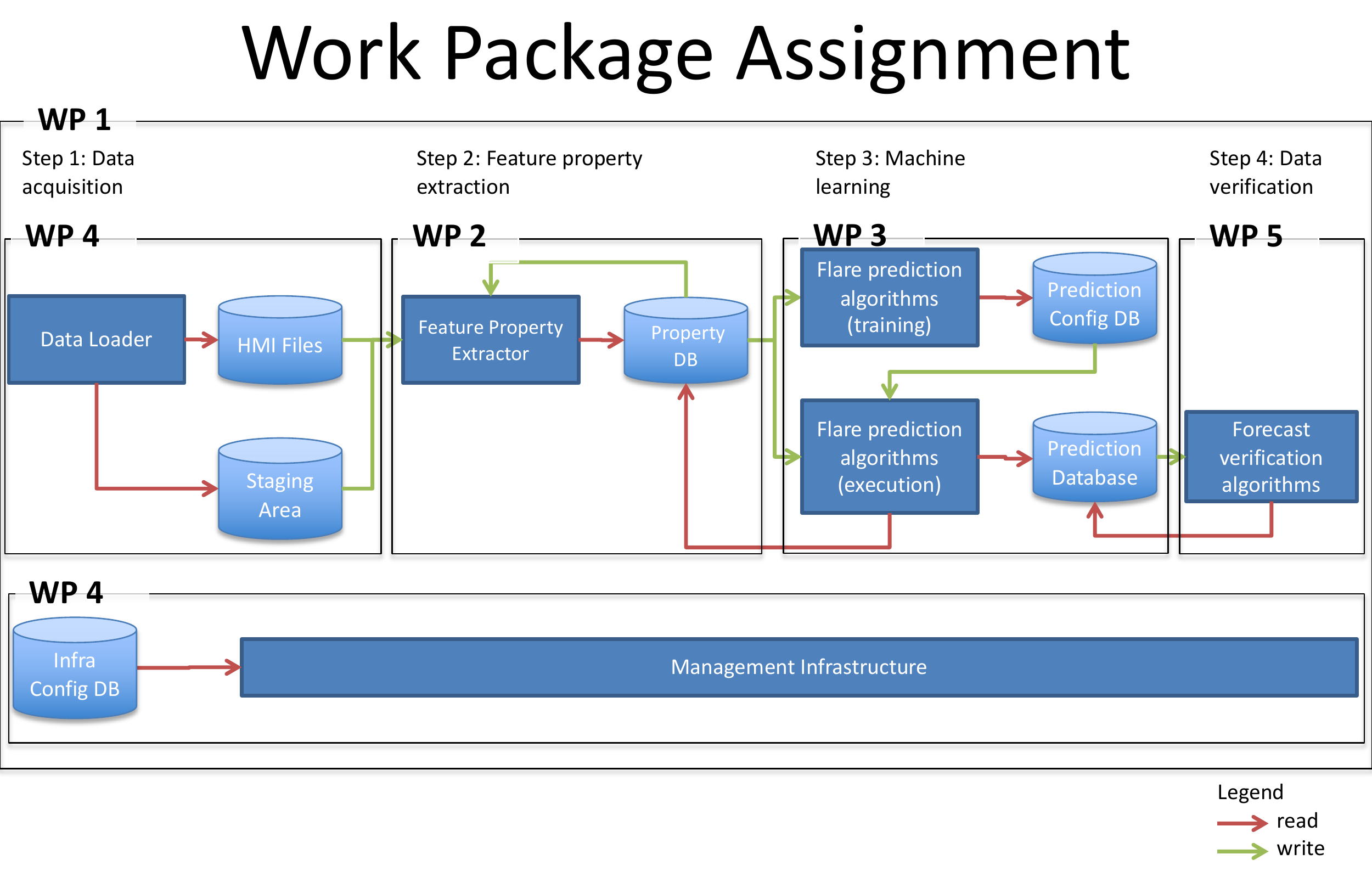}
    \caption{The FLARECAST architecture and WP Assignment. 
    Rectangles indicate exchangeable software components, such as algorithms and codes, while cylinders indicate data model components. 
    WPs 6 and 7 are not included here but are integral to the project and complement the overall efforts.}
    \label{fig:wp_assignment}
\end{figure}
FLARECAST embraced the technical architecture and methodology structure illustrated in Figure~\ref{fig:wp_assignment}. 
This was realized in a sequence of four procedural steps, namely, (1) [external] data acquisition, (2) property extraction, (3) machine learning-based prediction and (4) forecast verification. 
The overall project methodology was implemented in seven WPs, as follows: 
\begin{itemize}
    \item {\bf WP1:} Project Management.  
    \item {\bf WP2:} Active Region Properties as Predictors of Flaring Activity.  
    \item {\bf WP3:} Flare Prediction Algorithms. 
    \item {\bf WP4:} Data Storage and Processing Cloud.  
    \item {\bf WP5:} Data and Forecast Verification.  
    \item {\bf WP6:} Explorative Research. 
    \item {\bf WP7:} Dissemination. 
\end{itemize}
This work plan structure, together with the technical scheme of Figure~\ref{fig:wp_assignment}, show the design philosophy of the FLARECAST science and technology. 
The project relied on machine learning applied to data from 
HMI vector magnetograph \citep{scherrer12,schou_etal12} onboard the SDO mission \citep{pesnell_etal12}
in order to implement a technological service for flare forecasting with the ultimate goal of contributing to a data-driven understanding of the physical triggering mechanisms of flares. 
Rigorous forecast verification and dissemination of results have been crucial steps of this basic research effort.

\subsection{Science motivation}
\label{s2:sci}

The solar flare phenomenon pertains to explosive energy release in the low solar atmosphere that results in short-lasting (i.e., minutes to hours) enhancement of emission over virtually the entire electromagnetic spectrum (see, e.g., \citealt{benz_08} and \citealt{Fletcher_etal11} for comprehensive reviews). 
Major, highly-energetic flares (i.e., GOES M- and X-class events, in particular) are much less frequent 
\citep[in a distribution that is long known to be a power law -- see][]{drake71,rosner_vaiana78,crosby_etal93}
than minor flares (i.e, GOES C-class) and subflares\footnote{For a definition of NOAA/GOES flare classes, see \url{https://www.swpc.noaa.gov/phenomena/solar-flares-radio-blackouts.}}. 
In terms of mean occurrence frequency, or \textit{climatology} in the statistical language of forecasting problems \citep[e.g.,][]{barnes_etal16,leka_etal19a}, major flares fall under the category of \textit{rare} events, namely, events that are much more infrequent than the physical systems in which they appear (in this case, solar ARs). 
A typical 11-year sunspot cycle involves the appearance, evolution and fading of a few thousand NOAA-numbered ARs, yet Carrington's flare is considered a one-in-a-century (or even more rare) event. In other words, viewing a set of flaring ARs as a `positive' sample vs. a `negative' sample of non-flaring ones, the ratio of sample sizes is substantially different than one. 
Increasing the flare magnitude threshold between positive and negative samples only pushes this ratio to further extremes. 
For example, in Solar Cycle 23 only $\sim$1.8\% of ARs hosted at least one GOES X-class flare (an imbalance ratio of $\sim$0.0183), with a respective ratio of $\sim$0.005 for GOES $\geqslant$X5.0 flares. 
The climatology of the FLARECAST flare sample in the even weaker Solar Cycle 24 is 1 GOES C-class flare every $\sim 11$ hours, 1 M-class flare every $\sim 4.5$ days and 1 X-class flare every $\sim 67$ days; substantial variations obviously exist over different phases of the cycle.
Class imbalance in major flare prediction and other rare-event problems is a central concern \citep{woodcock_76,jollisteph_12,bloomfield12,bobra_couvidat15} for machine learning methods with proposed remedies including undersampling, oversampling and misclassification weighting \citep[e.g.,][]{longadge_etal13,ahmadzadeh_etal21}.

Flares at the top end of the observed energy distribution are viewed and treated as \textit{extreme} events, given their very significant disruptive ability on top of their scarcity. 
For diverse accounts of extreme events in physical systems one may review \citet{albe_etal06,meyers_11,sharma_etal12}, among other comprehensive works. 
These accounts also refer extensively to two intrinsic characteristics of extreme events: environmental complexity and difficulty in forecasting. 
%For flares, in particular, prediction difficulty due to stochasticity is further spurred by a long-standing notion that they are responses of nonlinear dynamical systems in a Self-Organized Critical State \citep[e.g.,][and references therein]{aschwanden_etal16}. 
It is a fact that forecasting solar flares is a pressing issue for space-faring nations, mainly for two reasons: first, because of flares' biological and technological repercussions and, second, because flares are a common element of the two other aspects of extreme space weather, CMEs and SEPs. Notwithstanding the lack of an early warning for flares, mentioned already, just
%Establishing a viable flare prediction would undoubtedly help make strides in the more general problem of space weather forecasting. However, there is no early warning time for solar flares; their hard X- or even $\gamma$-ray photons are already at Earth when observed by the GOES spacecraft. Only 
a cursory examination of some biological implications indicates that, say, a 500 keV $\gamma$-ray photon has a wavelength of $\sim$0.25 nm. 
This is much shorter than a DNA helix ($\sim$3.5 nm), meaning that such radiation acting on exposed humans engaging in extravehicular activities can lead to acute radiation sickness, if not being altogether fatal (see, for example, \citealt{freese_etal16}). 

The complexity, in terms of the multiscale (i.e., multifractal) behavior in the turbulent solar atmosphere (see, for example, \citealt{georgoulis_05}) undoubtedly adds to the difficulty of predicting solar flares. 
In particular, flares are long thought to be responses of nonlinear dynamical systems (solar ARs) in a SOC state. 
The SOC concept was initiated by seminal works on the topic \citep{lu_hamilton91,lu_etal93,vlahos_etal95} inspired by groundbreaking developments in theoretical physics (\citealt{bak_etal87,bak_etal88,kadanoff_etal89}; see also \citealt{bak_96} for an encompassing account). 
An account of apparent SOC manifestations in Astrophysics, including flares, is presented in relatively recent anniversary works \citep{aschwanden_etal16,mcateer_etal16}. 
However, a SOC evolution of flaring ARs would lead to an intrinsic stochasticity in flare occurrence; this translates to an intrinsically probabilistic forecasting relying by definition on small probabilities for major and extreme flares. Randomly driven SOC models explore precisely this stochasticity; hence they are incapable of predicting small- and mid-size flares. However, \citet{strugarek_charbonneau14} reported that a class of deterministically driven SOC models could be used for predictive purposes as they raise the memory of the system, thus exerting a partial control over flare waiting times;  however, this is an avenue of research yet to be sufficiently explored. Recent studies have also aimed to assess the frequency of a Carrington-level event in the framework of extreme value theory \citep{elvidge_angling18}. Rather than focusing exclusively on extreme events, the key scientific question and incentive behind the FLARECAST project was to determine to what degree the skill currently achieved in the forecasting of major flares could be advanced. 

Correlating solar flares with magnetic evolution  dates back several decades. One of the earliest,  pioneering accounts was that of \citet{howard_severny63}, who reported major magnetic field changes before and after a major (coined as 'great') flare. More systematic works were added in the 1980s with limited-resolution magnetograms \citep{krall_etal82,hagyard_etal84,zirin_liggett87} linking flares --and even repeated flaring-- to $\delta$-sunspot complexes and velocity shear, before the first semi-operational flare prediction schemes appeared \citep{mcintosh90,Zirin91}. In more recent years, however, the formation of long and intense magnetic PILs in the photospheric magnetic field of ARs was established as a feature of direct relevance to major flaring \citep[for comprehensive reviews, see][and references therein]{schrijver09,toriumi_wang19}. By 'intense', we mean PILs  exhibiting substantial amounts of magnetic flux and strong magnetic gradients due to the closely seated, opposite magnetic polarities and magnetic / velocity shear \citep[e.g.,][and references therein]{georgoulis_2019,patsourakos_etal20}.
Major flares occur as such PILs evolve and intensify, fueled by magnetic flux emergence and cancellation along them \citep{vanb_martens89,gibson_fan06,archontis_torok08}. Sometimes, an eruption including a major flare may occur in the absence of intense PILs, when the emerging flux reconnects with the pre-existing field or has enough magnetic twist to become unstable during  emergence. Such a exceptional case, giving rise to a GOES X3 flare, was reported by \citet{allen_moore04}. Regardless, the majority of X-class flares --the ones mostly affecting space weather conditions-- occur above evolving, intense photospheric PILs. 

\begin{figure}
    \centering
    \includegraphics[width=0.8\textwidth]{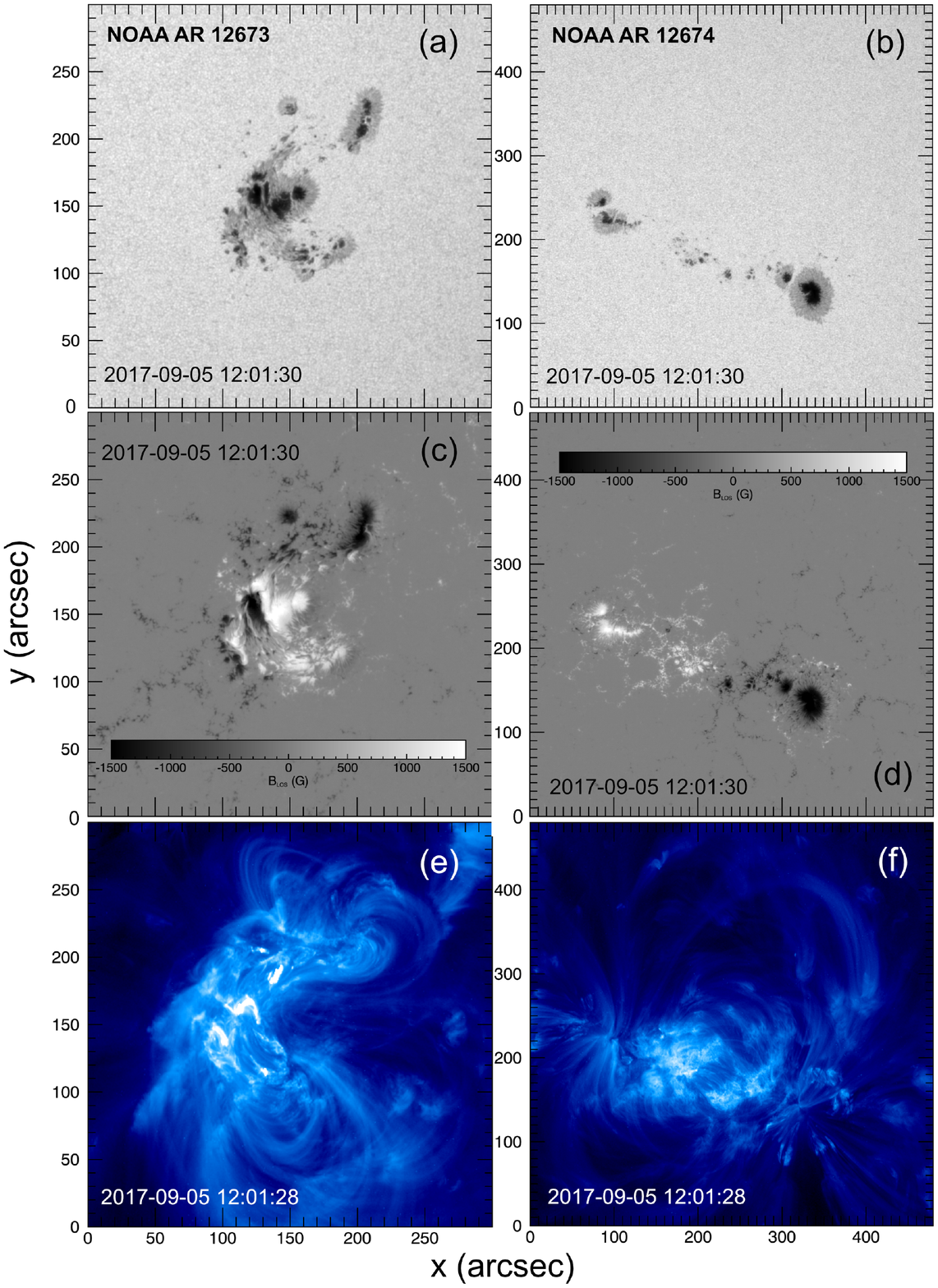}
    \caption{Examples of two solar ARs, in terms of their photospheric continuum intensity images (a,b), LOS magnetograms (c,d) and EUV coronal images at 19.3~nm (e, f). Both active regions were observed on 5 September 2017 at around 12:01 UT. Of them, NOAA AR 12673 (left column) was intensely eruptive, giving the largest eruptive flares since January 2005, while NOAA AR 12674 (right column) did not host any major flares. Images (a-d) have been acquired by HMI, while images (e-f) by AIA, both onboard the SDO mission.}
    \label{fig:ar_example}
\end{figure}
An example of ARs with and without strong PILs is given in Figure~\ref{fig:ar_example}, where the relatively flare-quiet (up to mid-C-class flares) NOAA AR  12674 is compared against the intensely flaring NOAA AR 12673. The latter in September 6 and 10 2017 gave the strongest ($\sim$X10) flares of solar cycle 24 \citep[e.g.,][]{yan_etal18}. Both active regions evolved simultaneously in the solar disk, located just a few hundred arcsec away from each other. While the photospheric compactness and conspicuous PILs are evident in NOAA AR 12673, NOAA AR 12674 is much more scattered. Coronal information in Figure~\ref{fig:ar_example} (sampled indicatively at 19.3~nm to showcase some structure) indeed shows significant complexity in terms of several bright kernels, complexity and apparent twist; significant non-potentiality, in brief \citep[e.g.,][]{schrijver_etal05}. Emission generally lacks bright kernels and does not show such non-potentiality in NOAA AR 12674. 

\begin{figure}
    \centering
    \includegraphics[width=0.85\textwidth]{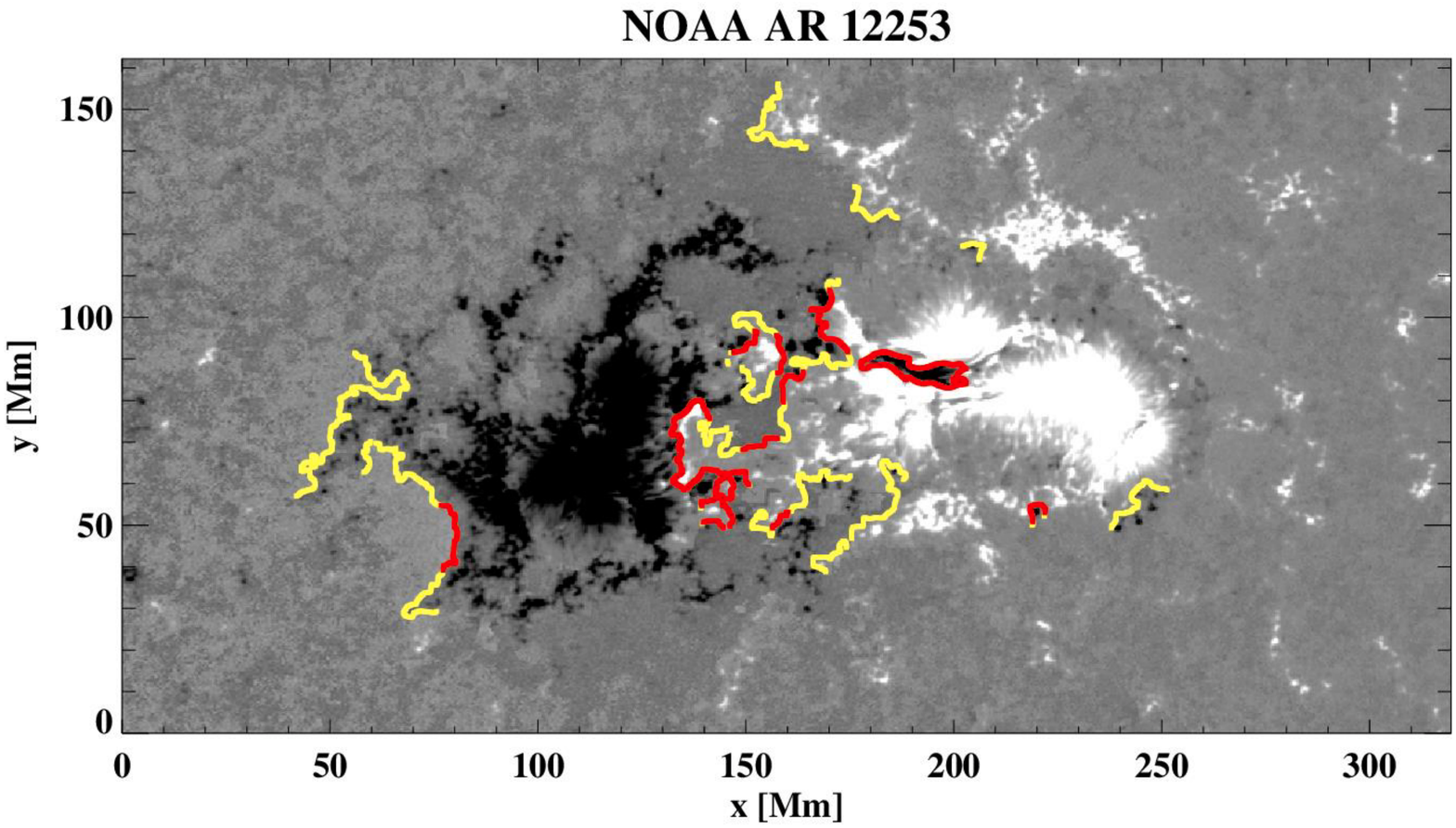}
    \caption{Automated PIL identification in a photospheric longitudinal magnetogram of NOAA 12253, observed by SDO/HMI on 2 January 2015. 
    The identified PILs in the region are further characterized as 'moderate' (yellow contours) and `strong' (red contours).
    The difference between 'strong' and 'moderate' PILs relies on the absolute value of the gradient of the vertical magnetic field ($>$40 G/pixel and $>$16 G/pixel, respectively) and the strength of the horizontal field ($>$120 G and $>$100 G, respectively). Parts of the PIL that are not highlighted in color are below one or both thresholds and are considered as 'weak'. The algorithm used is the FLARECAST PIL identification code, accessible as described in Appendix \ref{app1}.2.    
    }
    \label{fig:AR12253}
\end{figure}

Figure~\ref{fig:AR12253} further shows a PIL analysis on NOAA AR 12253, achieved in FLARECAST's framework. 
More information is given in Section~\ref{sec:NewPredictor}. 

\subsection{Data}
\label{data}

\subsubsection{Active region properties and flare predictors}
Flare forecasting relies almost entirely on statistical correlations between solar magnetic field data and observed flare characteristics. 
In this case, local (i.e., AR scale) photospheric magnetic fields are parameterized in order to identify and quantify patterns potentially associated with flares. 
Should a flare occur, flare detection and classification is primarily done using the GOES $0.1-0.8$~nm soft X-ray band. The peak photon flux in this band is historically used to determine the GOES flare class.
There are obviously other sophisticated ways to detect flares in EUV, X-ray, and optical wavelengths \citep[see][for an attempt in the framework of the HEK project]{martens_etal12}, but the GOES X-ray classification is the one used most widely by the space weather and flare forecasting communities. Although the GOES soft X-ray detectors are, in essence, spectral irradiance instruments without spatial resolution on the solar disk, observations by the GOES/SXI telescopes complement flare information with the source ARs, at least for the largest events. However, these spatial identifications are not error-free and verification of all GOES flare locations is a nontrivial task. Significant effort on identified flare locations was also put by \citet{hock_12} and, more recently, by \citet{angryk_etal20}. 

Quantitative active region photospheric properties, used as flare  predictors in FLARECAST's framework, were derived from the HMI SHARP data product \citep{hoeksema_etal14, bobra_etal14}. 
A FLARECAST processing pipeline was developed using the \texttt{hmi.sharp\_cea\_720s\_nrt} data series that contains AR maps of the LOS magnetic field component, $B_{\rm LOS}$, continuum intensity, $I_{c}$, and vector magnetic field components in the radial, $B_{r}$, co-latitude, $B_{\theta}$, and azimuthal, $B_{\phi}$, directions of the solar spherical coordinate system. 
These maps are produced at HMI's JSOC\footnote{\url{http://jsoc.stanford.edu/ajax/lookdata.html?ds=hmi.sharp\_720s\_nrt}} 
at a cadence of 12~min, while the NRT HMI stream assures that these data products are available 
within typically 4~hr of acquisition{\footnote{This time has been varying for different intervals of the SDO mission. With the current NRT data level, data are made available typically within 3 hours.}} \citep{hoeksema_etal14}. Another reason for using the NRT stream was that any operationally-oriented flare forecasting service has to rely on NRT data for training, testing, and validation of its method(s).
In a further attempt to replicate operational conditions to the greatest extent possible, FLARECAST did not restrict its input data to longitudinal areas close to the central solar meridian, but identically treated all data regardless of solar disk position (for regions very close to the solar limbs though, some selection criteria were imposed, as discussed below). 
This was decided in spite of a complete understanding that magnetic projection effects, foreshortening and noise near the solar limbs are substantial. 

In the FLARECAST processing pipeline, NRT SHARP maps are pre-processed before extracting any property. Pre-processing aims to: 
\begin{enumerate}[label=(\alph*),leftmargin=*]
    \item check for missing information in the FITS headers and restore it, if possible;
    \item examine for bad-quality or missing data (i.e., a NaN or a constant value);
    \item capture possible differences between WCS 
    \citep{thompson06} positions and header positions; 
    \item trim any part of the FOV containing off-disk pixels for ARs near the limbs.
\end{enumerate}
Magnetogram maps are flagged as null and no properties are calculated for HARP timestamps and dates with (i) bad-quality data, (ii) absolute differences between FITS header coordinates and WCS-calculated coordinates that are larger than five degrees or (iii) trimmed images that are less than 20 pixels in horizontal size. 

Active region properties extracted from the photospheric magnetic field in FLARECAST have all been previously tested for relevance in flare prediction. These properties are presented in Table~\ref{tab:FC_props}, itemized into 15 different groups. Each property group relates to either a certain property or a set of properties and gives rise to one or more flare predictors, respectively. These predictors, and groups thereof, quantify the state of the AR photosphere and corona and reflect (1) the entire region or SHARP FOV, (2) magnetic PILs, or (3) sunspots (as the areas with the strongest magnetic field). 
Most property groups characterise only the state of the photospheric magnetic field. 
Some, however, characterize the state of the AR corona, either by photospheric proxies or by means of current-free, potential magnetic field extrapolations (PFEs) or by differences between the photospheric field and that expected by a PFE.
%Table~\ref{tab:FC_props} displays the complete list of properties, their respective groups 
%(for detailed property definitions and calculations, refer to the respective literature notes therein) 
%and the number of individual quantities (hereafter called predictors) that they produce for use by the prediction algorithms. 
The two last columns of Table~\ref{tab:FC_props} indicate relevant published work, distinguishing between studies that were directly adopted in the course of the project from additional or previous studies pertinent to some predictors. 

It should be mentioned that, for many of the properties in Table~\ref{tab:FC_props}, estimating uncertainties is nontrivial. For the cases uncertainties exist, mostly in terms of $\sigma$-values in fits, these values are provided in the property database. This said, prediction algorithms (Section \ref{s2:ml}) currently do not utilize predictor uncertainties. Uncertainties in verification metrics are provided, however, as per the sensitivity analysis of Section \ref{s:uncert}. Evaluating the impact of predictor uncertainties in multi-parametric machine learning prediction schemes could be a meaningful future step.

Up to 209 predictors are calculated from each SHARP observation at different cadence, namely 1~hr, 3~hr, 6~hr, 12~hr, and 24~hr. The full 12-min cadence of HMI SHARPs was exploited only in a very limited number of cases, due to the immense computational time it required. This computational expense dictated the different cadence as a necessity. Clearly, FLARECAST did not restrict to the 25 scalar properties included in the SHARP predictor metadata. It did, however, replicate the calculations of these predictors to validate them and consider additional image statistics (i.e., higher-order moments). Instructions on how to access these and other data and information are detailed in Appendix~\ref{app1}.

\begin{landscape}
\begin{table}
\scriptsize
\caption{The complete list of FLARECAST properties, separated in property groups and accounting for a total of 209 predictors. The vast majority (197) of them correspond to the AR magnetic field, while the rest include information from the NOAA SWPC Catalogues and from photospheric intensity images.}
\label{tab:FC_props}
\begin{tabular}{l l l l l l}    
\hline
\textbf{Data Source} 		& \textbf{Property Group}  									& \textbf{No. of Predictors} & \textbf{Relevant Predictor} & \textbf{Adapted from} & \textbf{Related references}	\\
\hline
SWPC  	            & Solar Region Summary (SRS) properties   			&                     2 & McIntosh and Hale classes; & \citet{mccloskey_etal16} & \citet{mcintosh90}\\
                    &                                			        & 3 & Number, area and longitudinal extend of sunspots & & \textcolor{red}{\citet{lee_etal12}}\\
Catalogues          & GOES soft X-ray flare events$^{\mathrm{a},\mathrm{b}}$            	& 4 & Flare magnitude, start, peak, and end times & & \\ 
\hline
Surface-normal 	    & Effective connected magnetic field strength$^{\mathrm{c}}$ & 1 & $B_{\mathrm{eff}}$ & \citet{georgoulis07}, & \\
component           &                                                   & & & \citet{georgoulis2010,Georgoulis_2013} & \\
(radial and         & Fractal and Multifractal parameters          		& 1 & Fractal dimension & \citet{Conlon2008} & \citet{Abramenko2003},\\ 
line-of-sight)      &                                                   & 1 & Generalized correlation dimension &  & \citet{Abramenko2005},\\ 
magnetograms        &                                                   & 2 & Holder exponent; Hausdorff dimension & & \citet{Al-Ghraibah2015}\\  
                    &                                                   & 2 & Structure function's Inertial range index & & \\
                    & Fourier and Wavelet power spectral indices        & 2 & Power-law exponent & \citet{2008SoPh..248..311H}, & \\
& & & & \citet{2015SoPh..290..335G}& \\
                    & Decay index (DI)$^{\mathrm{d}}$                    & 8 & Mean DI over PIL segments; & \citet{Liu2008}, & \\ 	
                    &                                                   & & Height of DI; Ratio of PIL length to DI height & \citet{Zuccarello2014} & \\
                    & Magnetic PIL properties & 5 & Sum of PIL segments, Longest PIL segment & \citet{2010ApJ...723..634M} & \\
                    &                                			& 1 & $R$ value & \citet{2007ApJ...655L.117S} & \\
                    &                                                   & 1 & $WL_{sg}$ & \citet{2012ApJ...757...32F} & \\
                    & 3D magnetic null points$^{\mathrm{d}}$             & 6 & Number of null points in different height ranges & \citet{Haynes2007} & \citet{Pontin2013} \\
                    &											
                    & & (from 2 to 100 Mm above photosphere) & & \citet{Barnes2006}\\
                    & Ising Energy$^{\mathrm{c}}$                                & 6 & Original and partitioned Ising energy & \citet{ahmed10} & \citet{kontogiannis18}\\
                    & Magnetic energy and helicity                  	& 11 & Poynting flux and magnetic helicity flux proxies & \citet{Park2010}, & \\
                    &													& & & \citet{park12} & \\
\hline
Full-Vector         & SHARP properties$^{\mathrm{e}}$                     & 100 & Horizontal gradient of $B$ components; & \citet{bobra_etal14} & \\
magnetograms        &                                                   & & Shear angle; Unsigned vertical current; & (validated) & \\
& & &higher-order moments of timeseries & \citet{leka_barnes03,leka_barnes07}& \\
		            & Magnetic energy and helicity          		& 22 & Poynting flux and magnetic helicity flux & \citet{Kusano2002} & \citet{berger_field_1984}, \\
                    &                                            	 	& & & & \citet{Welsch2009} \\
                    & Non-neutralized Currents                     		& 6 & Total non-neutralized current & \citet{geo_titov_mikic12} & \citet{kontogiannis17} \\
                    & Flows around PIL             		            & 22 & Speed of diverging/converging/shear flows & \citet{park18}, & \citet{Deng2006},\\
& & & & & \citet{Wang2014}\\
\hline
Intensity Images$^{\mathrm{f}}$ & Magnetic field gradient                        	& 3 & Total horizontal magnetic gradient & \citet{korsos14} & \citet{kontogiannis18} \\
\hline
\end{tabular}\\
$^{\mathrm{a}}$ AR coronal information.\\ 
$^{\mathrm{b}}$ The project uses the flare attributes as properties. At least one flare of desired magnitude in the forecast window signifies a positive instance.\\
$^{\mathrm{c}}$ Photospheric proxy for coronal information.\\
$^{\mathrm{d}}$ Uses PFEs.\\
$^{\mathrm{e}}$ Most SHARP parameters correspond to mean values. In the FLARECAST pipeline, other relevant magnetogram-related parameters are also used as predictors; these are maximum and minimum values, median, standard deviation, kurtosis, and skewness.\\
$^{\mathrm{f}}$ Used in conjunction with magnetograms to calculate the sum of the magnetic field gradients between all possible opposite-polarity umbrae pairs.
\end{table}
\end{landscape}

\begin{figure}
    \centering
    \includegraphics[height=0.8\textheight]{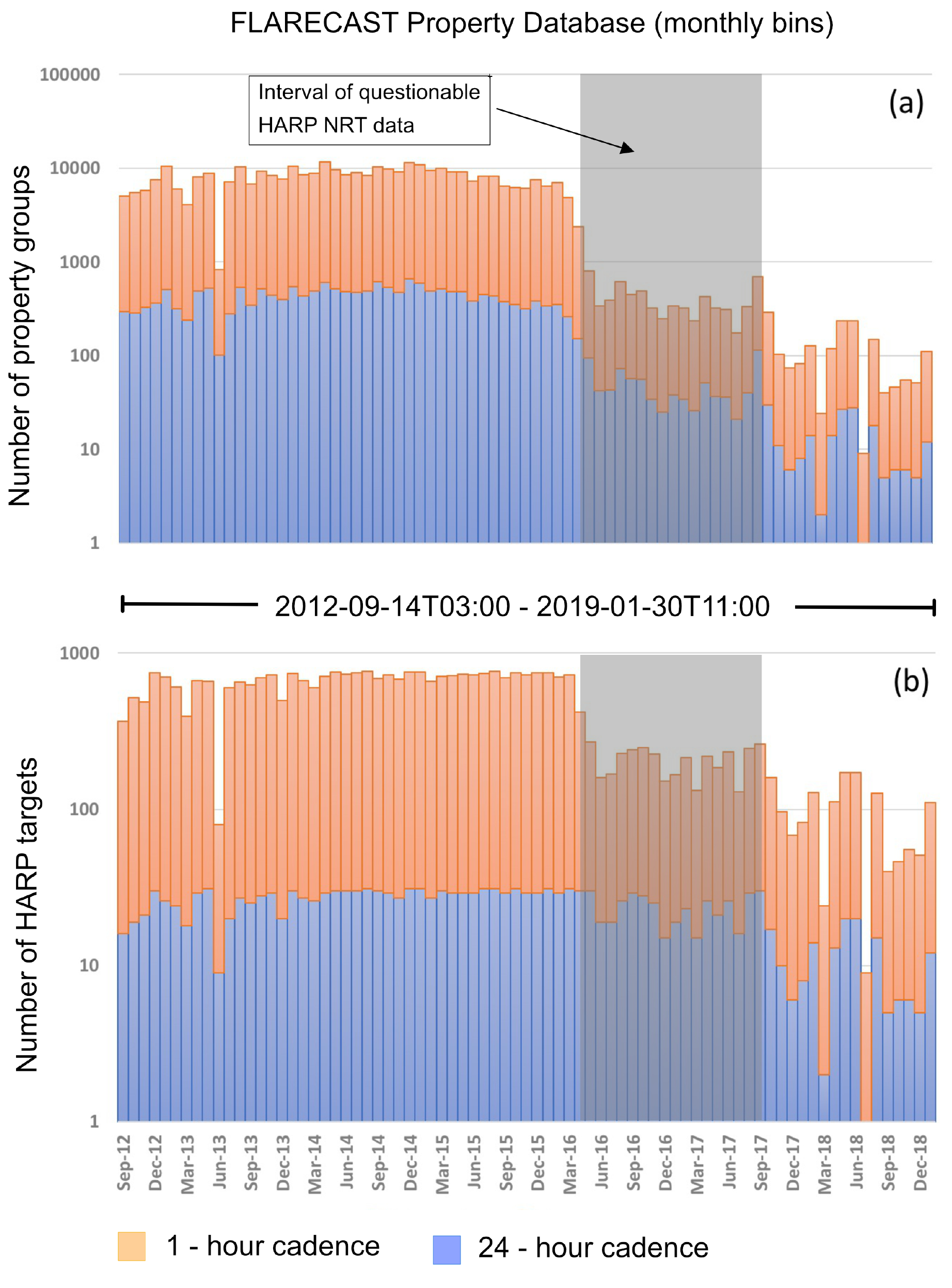}
    \caption{Monthly numbers of (a) calculated property groups and (b) processed NRT SHARP timestamps between 14 September 2012 and 30 January 2019. Property group numbers are shown at highest (1-hr; orange bars) and at lowest (24-h; blue bars) cadence. 
    The shaded interval between 13 April 2016 and 8 September 2017 corresponds to a period of questionable quality for the HMI NRT SHARP data (see text for details). 
    Histograms represent a total of 32,098 processed SHARP timestamps, leading to millions of predictor values, since each timestamp could potentially provide up to 209 predictors.}% with 336,166 property groups calculated.}
    \label{fig:DB_hist}
\end{figure}

Figure~\ref{fig:DB_hist} displays histograms of the monthly number of property groups and SHARP timestamps analyzed over the period September 2012 to January 2019, for the finest (1~hr) and the coarsest (24~hr) cadences considered. Numbers corresponding to 12-minute cadence are not shown due to their lack of statistical significance.
The progression of Solar Cycle 24 can be roughly assessed, with significantly higher numbers until early 2016, when the declining phase gave way to the latest solar minimum. 
Thereafter, the numbers of both calculated property groups and processed SHARP timestamps gradually decline as eligible SHARPs in the NRT stream become fewer and fewer. The shaded interval between 13 April 2016 and 1 September 2017 corresponds to a time of potentially problematic NRT data due to a transient misalignment between the two HMI cameras providing filtergrams for the generation of the full Stokes vector\footnote{See \url{https://solarnews.nso.edu/20170901/\#section_hoeksema} for more information.}. Only definitive vector magnetogram data were reprocessed --- NRTs were not --- while full-disk LOS magnetograms at 45s cadence were not affected. Artifacts in this case intensify with increasing central meridian distance.

The property data set shown in Figure~\ref{fig:DB_hist} is, to our knowledge and understanding, one of the largest and most diverse ever assembled for flare prediction, offering unique opportunities for statistical and physics-based studies to interested teams worldwide. In \citet{Guerra2018}, a sub-group of six predictors 
%($\alpha$, $B_{\rm eff}$, $L_{\rm tot}$, $R$, $E_{\rm Ising}$, $(L/h_{\rm min})_{\rm max}$) 
were studied to determine the effects of using the LOS field ($B_{\rm LOS}$) vs. the surface-radial field ($B_{r}$) for their calculation and subsequent use in flare forecasting. It was determined that both LOS and radial field components can be advantageous in different circumstances. The project hence decided to use both property versions to facilitate any independent information these two versions could furnish. 

\subsubsection{Flare association}

The NOAA/SWPC flare event list data include the universal times (UT) of their start (i.e., onset), peak, and end, along with the flare magnitude, all obtained by the GOES spacecraft $0.1-0.8$~nm soft X-ray channel. Where possible, source ARs are identified from the daily NOAA/SWPC flare event lists, with their morphological properties extracted from the daily NOAA/SWPC SRS reports. NOAA ARs, flaring or not, are linked to the HMI SHARPs and the FLARECAST property database includes information on all ARs located within each SHARP FOV. 

The process below is followed for each SHARP during its solar disk passage:
\begin{enumerate}
    \item SHARPs are first checked to determine whether their FOV contains the time-advanced centroid locations of NOAA-numbered ARs. 
    This makes use of the closest SRS report before the SHARP observation, with solar differential rotation taken into account. %in the coordinate comparison.
    \item If any NOAA ARs are assigned to the SHARP, the SWPC flare event list is searched for those NOAA numbers and related flares are associated to all of that SHARP's property database entries. 
    \item For X-ray flares with no reported NOAA number, locations of co-temporal 
    flares observed in ground-based H$\alpha$ images (also from the NOAA/SWPC flare event list) are used to determine if the multi-wavelength flare event occurred within the SHARP's FOV (again taking solar differential rotation into account). Positive H$\alpha$ flare association results in the inclusion of the respective X-ray flare. A very small fraction of flares of GOES class C and above, of the order 0.1\%, seems to miss both criteria above over the analysis period.
\end{enumerate}
After all SHARP-associated flares are identified, we extract the following properties for each flare:
\begin{itemize}
    \item $F_{M}$: GOES peak magnitude (e.g., M1.3)
    \item $\tau_{s}$: Time difference (in seconds) between the SHARP observation time $T_{0}$ and the reported flare start time $T_{s}$ (i.e., $\tau_s=T_{s} - T_{0}$)
    \item $\tau_{p}$: Time difference (in seconds) between $T_{0}$ and the reported flare peak time $T_{p}$ (i.e., $\tau_p=T_{p} - T_{0}$)
    \item $\tau_{e}$: Time difference (in seconds) between $T_{0}$ and the reported flare end time $T_{e}$ (i.e., $\tau_e=T_{e} - T_{0}$).
\end{itemize}
%Properties extracted from SHARP-associated flares are then assigned to the same property group (SRS properties and GOES soft X-ray flare events in Table~\ref{tab:FC_props}). 

\begin{figure}[t]
    \centering
    \includegraphics[width=0.93\textwidth]{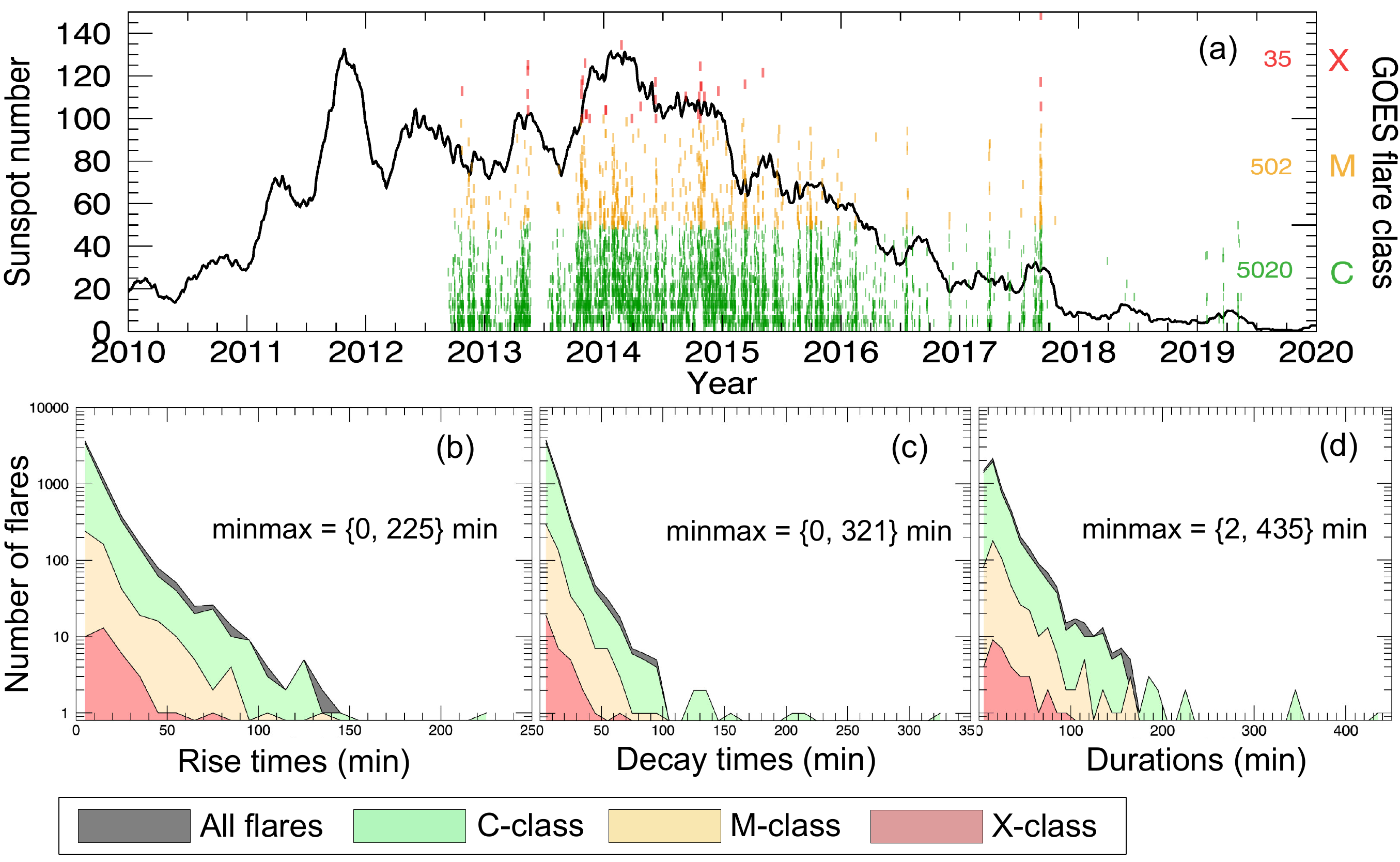}
    \caption{Attributes of 5,557 SHARP-associated flares over September 2012 to May 2019. %(5,526 of them corresponding to the interval of Fig.~\ref{fig:DB_hist})
    (a) Flare onset times, in terms of GOES classes X (red), M (orange) and C (green) and respective sub-classes (tick marks; right ordinate), plotted over the 90-day averaged sunspot number over Solar Cycle 24 (left ordinate; Source: Sunspot Index and Long-Term Solar Observations [SILSO], Royal Observatory of Belgium). The numbers of flares for each GOES class are also given. The bottom line shows waterfall diagrams of the distributions of rise time (b), decay time (c) and duration (d). Extrema for each distributions are shown in  each plot. The 'All flares' diagram corresponds to the sum of flare numbers in each bin. We use fixed, 10-minute bin sizes.}
    %Flares of GOES class C and above, C only, M only and X only are shown by gray, green, orange and reddish columns, respectively.}
    \label{fig:Flares_hist}
\end{figure}

Figure~\ref{fig:Flares_hist} provides a snapshot of FLARECAST's flare data set and displays the temporal distribution of different flare classes for Solar Cycle 24 (Fig.~\ref{fig:Flares_hist}a), along with their rise time (Fig.~\ref{fig:Flares_hist}b), decay time (Fig.~\ref{fig:Flares_hist}c) and duration (Fig.~\ref{fig:Flares_hist}d) for approximately the same time coverage as that of the property database statistics in Figure~\ref{fig:DB_hist}. Rise/decay times and durations have been provided by NOAA.

We consider as eligible flares only those of GOES C-class and above (i.e., $\geqslant$C1.0) to make sure that as many as possible are included, unobscured by an often elevated solar soft X-ray background. 
We acknowledge, though, that even a C1.0 flare threshold might result in loss of some flares in cases of intensely high solar activity. 
Others may be lost due to lack of, or an erroneous, location information.

%Distributions of flare duration ($\tau_{e} - \tau_{s}$), rising time ($\tau_{p} - \tau_{s}$), and decaying time ($\tau_{e} - \tau_{p}$) are shown in Figure~\ref{fig:Flares_hist}(Left, Middle, and Right), in minutes. Rising and decaying time correspond the time scales of the flare impulsive and cooling phases. Duration time corresponds to the sum of rising and decaying times. It is observed in Figure~\ref{fig:Flares_hist}Left and middle that rising and decaying times for all flare classes have maximum occurrence for the lowest bin $\lesssim$ 10 min; then the occurrence values decrease as flare duration time increases. Maximum values observed in these distributions are τs = 100 min, τs = 150 min, and τs = 200 min for X-, M-, and C-class flares respectively. On the other hand, rising and decaying times (Figure 18, top) show comparable ranges of values ($0–150$~min) and comparable decreasing of the occurrence with increasing time values. 

\subsection{Methods: machine learning}
\label{s2:ml}

The FLARECAST computational component fully relied on machine learning, that is, on prediction algorithms that utilize an automatic learning step based on either labeled or unlabeled input data. The conceptual core of each machine learning technique lies on  the modality of this learning step. Labeled data are characterized by a tag including one or more property-specific labels (e.g., flaring or non-flaring), whereas unlabeled data are free of such tags.

We distinguish between two broad categories of relevant machine learning methods: 
\begin{itemize}
    \item \emph{Unsupervised methods}, that are free to infer the data structure from the data themselves and realize the learning task in a fully data-driven manner. Such methods train on  unlabeled data.
    \item \emph{Supervised methods}, that perform an unknown input-output mapping from known input-output samples. 
    Sparsity enhancing techniques in these methods enable a quantitative ranking of properties (predictors) that contribute most to the achieved prediction. Supervised methods nominally train on labeled data.
\end{itemize}
The project's machine learning inventory comprises three (3) unsupervised and eleven (11) supervised machine learning methods, listed in Table~\ref{tab:ML-methods}.
We chose to implement a diverse array of methods because we identified a need to experimentally test the strengths and weaknesses of each method, along with their key differences and predictive capacity. Identifying the state-of-the-art in machine learning methods flows directly from FLARECAST's main goal (Section \ref{sec:intro}).

\begin{table}[t]
    \centering
    \caption{FLARECAST unsupervised and supervised machine learning methods (final release).}
    \begin{tabular}{ll}
    \hline
    \textbf{Unsupervised methods} & -- K-means\\
                         & -- Fuzzy C-means\\
                         & -- Possibilistic C-means\\                     
    \hline
    \textbf{Supervised methods} & -- LASSO\\
                       & -- Hybrid LASSO\\
                       & -- Elastic net\\
                       & -- Logit\\
                       & -- Hybrid Logit\\
                       & -- Random forests (RFs)\\
                       & -- Multi-layer perceptron (MLP)\\
                       & -- Recurrent neural network (RNN)\\
                       & -- Support vector machine (SVM)\\
                       & -- Garson's method\\
                       & -- Olden's method\\ 
    \hline
    \end{tabular}
    \label{tab:ML-methods}
\end{table}

\subsubsection{Unsupervised methods}
FLARECAST proposes three unsupervised (i.e., clustering) methods for the automatic classification of sets of unlabeled AR properties. To briefly explain how these three algorithms work, we denote as $X = \{x_k~~|~~x_k \in \mathbb{R}^d~~~k=1,\ldots,n \}$ a set of unlabeled samples $x_k = (x_{1k},\ldots,x_{dk})$, where $x_{ik}$ is the $i$-th property of the $k$-th sample $x_k$, $d$ is the dimension of the property space $\mathbb{R}^d$ and  $Y=\{y_j~~|~~y_j \in \mathbb{R}^d~~,~~j=1,\ldots,c\}$ is the set of the $c$ centers of the clusters to determine. The algorithms' objective is to minimize a certain functional with respect to the cluster centers. The functionals of the K-means \citep{anderberg2014cluster} and Fuzzy C-means \citep{bezdek2013pattern} algorithms are given by,
\begin{equation}\label{K-means}
J(X,Y,U) = \sum_{k=1}^n \sum_{j=1}^c u_{jk} \delta_{jk}^{2}\ ,
\end{equation}
and,
\begin{equation}\label{FCM}
J_m(X,Y,U) = \sum_{k=1}^n \sum_{j=1}^c (u_{jk})^m \delta_{jk}^{2}\ ,
\end{equation}
respectively. 
In both equations above, $U=[u_{jk}]$ is the $n \times c$ membership matrix whose entries represent the memberships of the $k$-th sample to the $j$-th cluster and $\delta_{jk}$ is the distance of the $k$-th sample from the $j$-th cluster center.
In the case of the K-means algorithm, the memberships are binary values, while in the case of the Fuzzy C-means algorithm they are real numbers $\in [0,1]$, representing the membership probability. 
In the latter case, there is also a `fuzzifier' parameter $m$. 

The third clustering method is Possibilistic C-Means \citep{krishnapuram1996possibilistic,massone2006possibilistic}. 
This is an elaborate development of Fuzzy C-Means, in which each sample can, in principle, belong simultaneously to several clusters with different degrees of membership. 
In this case the cost function to minimize is given by,
\begin{equation}\label{PCM}
J_m(X,Y,U) = \sum_{k=1}^{n} \sum_{j=1}^{c} (u_{jk}^m \delta_{jk}^{2}) + \sum_{j=1}^{c} \eta_j \sum_{k=1}^{n} (1-u_{jk})^m\ ,
\end{equation}
where the entries of the membership matrix satisfy the constraint $\max_j u_{jk} >0~~\forall k$ and, for each $j$, the regularization parameter $\eta_j$ depends on the average size and shape of the $j$-th cluster.

\subsubsection{Supervised methods}
Most FLARECAST methods are supervised, in line with contemporary applications of machine learning to flare prediction \citep[e.g.,][]{ahmed13, benvenuto2018hybrid, bobra_couvidat15,Florios:18}. A detailed description of these methods is beyond the scope of this work.
The latest release of the project's platform contains a standard MLP trained by means of Error-Back-Propagation and two RNNs that allow feedback loops in the feed-forward architecture. 
This modification is realized by means of both an Elman neural network \citep{elman1990finding}, in which any number of context nodes is permitted, and a Jordan neural network \citep{jordan1997serial}, in which the number of context nodes is constrained to coincide with the number of output nodes.

In a more recent development, regularization neural networks \citep{evgeniou2000regularization} enable the connection between training and generalization via the minimization of functionals such as the following, 
\begin{equation}\label{regularization-network}
V(y_i,f(x_i)) + \lambda \|f\|_{\mathcal{F}} \rightarrow minimum~,
\end{equation}
where $\{(x_i,y_i)\}_{i=1}^{N}$ represents the training set made of $N$ property-label pairs, $V(\cdot,\cdot)$ is the loss function that measures the price paid for the inaccuracy of predicting $y_i$ with $f(x_i)$, and $\lambda$ is the regularization parameter, which realizes the trade-off between fitting over the training set and generalization. 

SVM for regression \citep{scholkopf2001learning} is one of the standard regularization networks implemented in FLARECAST. 
In this case, $V(\cdot,\cdot)$ is a standard quadratic loss function and ${\mathcal{F}}$ is a Reproducing Kernel Hilbert Space \citep{devito2004some}, in which four different kernel types can be selected (i.e., linear, polynomial, radial basis function, and sigmoidal). 
The FLARECAST platform also contains a SVM for classification that uses the hinge loss function, namely the one thought most appropriate for classification \citep{rosasco2004loss}. 
Two sparsity enhancing regularization methods were also implemented, in which the number of features that effectively contribute to the generalization is constrained to the smallest possible. 
This is achieved by minimizing the $l1$ norm of the feature vector (i.e., the sum of the absolute values of its components), which is further achieved by means of two different approaches: penalized logistic regression \citep{wu2009genome}, in which the loss function realizes the Bernoulli distribution for the labels, and LASSO \citep{yuan2006model}, in which the loss function is quadratic. 
FLARECAST also includes a hybrid version of penalized logistic regression and LASSO, in which the regression outcome is partitioned by means of a Fuzzy C-Means scheme, without focusing on optimizing a specific skill score \citep{benvenuto2018hybrid}. 
The platform contains a further generalization, namely an elastic net \citep{zou2005regularization} algorithm, in which the minimization functional contains two penalty terms ($l1$ and $l2$) with two different regularization parameters optimized by cross validation. 

Ensemble learning \citep{dietterich2000ensemble} is another supervised approach that uses a combination of different learning models to increase the classification accuracy. 
In this framework, FLARECAST offers a RF algorithm \citep{breiman2001random}, which works as a large collection of de-correlated decision trees. 
Given a training set of samples made of properties and corresponding labels, a decision tree recursively splits the training samples into subsets based on the value of a single property. 
Each split corresponds to a node in the tree and the task is to separate records in the training set that have different characteristics. We follow the implementation described in \citet{2002LiawWiener}, by splitting the tree until every subset is pure (i.e., all samples in the subset belong to the same class).
In this way all terminal nodes (i.e., the leaves) are assigned a unique class label. 
Once the decision tree has been constructed, classifying a test record is achieved by starting from the root node, applying the test condition to the record and following the appropriate branch based on the outcome of the test. 
This can lead either to another internal node, for which a new test condition is applied, or to a leaf node. 
If a leaf node is reached, the label associated with it is assigned to the record. 
In the RF approach, the training set is randomly divided into a fixed number of subsets and for each subset a decision tree is built. 
New, incoming unlabeled samples are classified by aggregating the predictions of the decision trees via a majority vote procedure.

Details on how to obtain the FLARECAST machine learning algorithms can be found in Appendix~\ref{app1}. 
We note in passing that the landscape of machine learning methods is continuously and rapidly evolving, with new algorithms constantly introduced. 
Our approach, as explained in Section~\ref{s2.3.3.} below, is an implementation based on modularity in such a way that incoming machine learning methods can be easily integrated in the FLARECAST platform. 

\subsection{Technology}
\label{s2:tech}
FLARECAST relies on a broad selection of different technologies, including hardware for storage and computing as well as software for data handling, infrastructure management and computation. 
The software was designed in a way that is hardware independent and can be installed on numerous platforms. All code developed during FLARECAST has been published under an open source license and is freely available. Code acquisition and license information are detailed in Appendix~\ref{app1}. 

\subsubsection{Computing hardware}
% FLARECAST infrastructure and technology used? from D4.6?
A computing server dedicated to FLARECAST % details about hardware?
has been integrated into the MEDOC computing infrastructure\footnote{{\url{https://idoc.ias.u-psud.fr/MEDOC}}} and hosts the production version of the FLARECAST pipeline.
Queries on this server for SDO/HMI data are made directly into the MEDOC database tables, and SDO/HMI files are accessed locally.
This allows efficient runs of the FLARECAST property extraction algorithms.

% data acquisition
\subsubsection{Data storage hardware}
The main FLARECAST data volume is provided by eight SDO/HMI data series at 12~min cadence that have been downloaded using the SDO NetDRMS and archived as part of the MEDOC solar physics data archive.
Besides the one used primarily for the forecasting tasks (\texttt{hmi.sharp\_cea\_720s\_nrt}), downloaded series included 
\texttt{hmi.m\_720s}, \texttt{hmi.m\_720s\_nrt}, \texttt{hmi.sharp\_720s},
\texttt{hmi.sharp\_720s\_nrt}, 
\texttt{hmi.sharp\_cea\_720s}, \texttt{hmi.ic\_720s} and 
\texttt{hmi.ic\_nolimbdark\_720s\_nrt}. Not all of them were finally used by the Consortium, although definitive SHARP data and continuum images were used for tasks akin to the explorative science component (Section \ref{sec:WP6}). 
NRT data series were downloaded no later than 1~hr after they were made available by HMI's JSOC. 
This overall delay could be further reduced, if necessary, by more frequent download requests.
% Other source data for FLARECAST are...

\subsubsection{FLARECAST software architecture}
\label{s2.3.3.}
The FLARECAST architecture design was driven by the needs for modularity, portability and ability to perform and accommodate different algorithms written in various programming languages. 
This philosophy is best described by the top-level diagram of Figure~\ref{fig:wp_assignment} in Section~\ref{s2}. 
The diagram reflects the four processing steps of the FLARECAST pipeline:
\begin{itemize}[leftmargin=*]
	\item {\bf{Step 1:}} Acquisition and transformation of data from multiple sources (SDO/HMI \& NOAA SWPC -- Section \ref{data}).
	\item {\bf{Step 2:}} Extraction of properties from the data by several algorithms (Section \ref{data}).
	\item {\bf{Step 3:}} Prediction through implementation of several machine learning algorithms (Section \ref{s2:ml}).
	\item {\bf{Step 4:}} Verification of the generated forecast data products (Section \ref{s4}).
\end{itemize}
The underlying management infrastructure controls and monitors these algorithms. 
%More verbose list
%\begin{enumerate}[label=\bfseries Step \arabic*:,leftmargin=*]
%	\item Import data from remote archives, such as GOES flare lists from NOAA Space Weather Prediction Center (SWPC) and magnetograms from the Helioseismic and Magnetic Imager (HMI) onboard the Solar Dynamics Observatory (SDO).
%	\item Extract features from the loaded data. This includes the parsing of external catalogue data into the FLARECAST data model as well as the extraction of features from imaging observations.
%	\item Train and execute prediction algorithms. This step covers both the training and the execution phases of the machine learning algorithms, as well as the execution of simpler statistical (Poisson) and discriminant analysis algorithms.
%	\item Verify the results from the previous step. This step relies on a large amount of prediction data for proper performance verification and benchmarking. Thus it is the most I/O critical part of the infrastructure.
%\end{enumerate}
The layout of the FLARECAST architecture simplified the project management as the individual components are under the responsibility of dedicated WPs. 
The relation between components and WPs is shown by the black rectangles in Figure~\ref{fig:wp_assignment}. 
The software infrastructure is discussed in detail in Section~\ref{s3}.
% Marco: not sure if we should keep this section. If not I would replace figure wp_assignment.pdf by architecture.pdf%

\subsubsection{Software components}
The software components, represented by the blue rectangles in  Figure~\ref{fig:wp_assignment}, fulfill specific tasks such as data loading, property extraction, machine learning, or verification. Generally, a software component contains several algorithms and each algorithm implements a specific variant of a task. Each property extraction or machine learning method, for example, has its own dedicated algorithm. 

The management infrastructure orchestrates the execution of the FLARECAST workflow. It launches the extraction algorithms as soon as new observational data arrives. After they are finished, the management infrastructure triggers the flare prediction algorithms.

The management infrastructure is implemented as a collection of containers (see Section~\ref{sec:dc} below). 
In automated mode, a set of repeatedly executed computer scripts (i.e., cron jobs) regularly check for new data and start the algorithms, if necessary. Additionally, a small web application allows users with administrator privileges to manually trigger the execution of algorithms.

\subsubsection{Data components}
The blue database symbols in Figure~\ref{fig:wp_assignment} denote individual parts of the FLARECAST data model, hereafter referred to as data components. 
Within the workflow, every software component (i.e., every algorithm) reads from several and writes to precisely one data component. Like this, the software components are decoupled from each other and only have to provide a control interface (start, stop) for the management infrastructure. The management infrastructure connects to a dedicated data component for configuration and data logging.

\begin{figure}
	\centering
	\includegraphics[width=\textwidth]{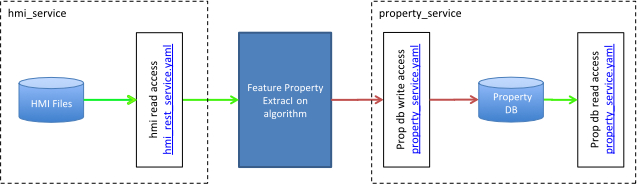}
	\caption{A data component (dashed box) consists of a database (blue cylinder) and a well-defined API (white box). 
	An algorithm (blue rectangle) uses this API to read data (green arrow) and another API to write data (red arrow) into a separate data component.}	
	\label{fig:data_access}
\end{figure}

Each data component defines a generic interface to read, create, update, and delete data. 
The read methods include a query language for simple queries. 
Figure~\ref{fig:data_access} illustrates this process for an example property extraction algorithm. 

\subsubsection{Docker containers}
\label{sec:dc}
\begin{figure}
    \centering
    \includegraphics[width=\textwidth]{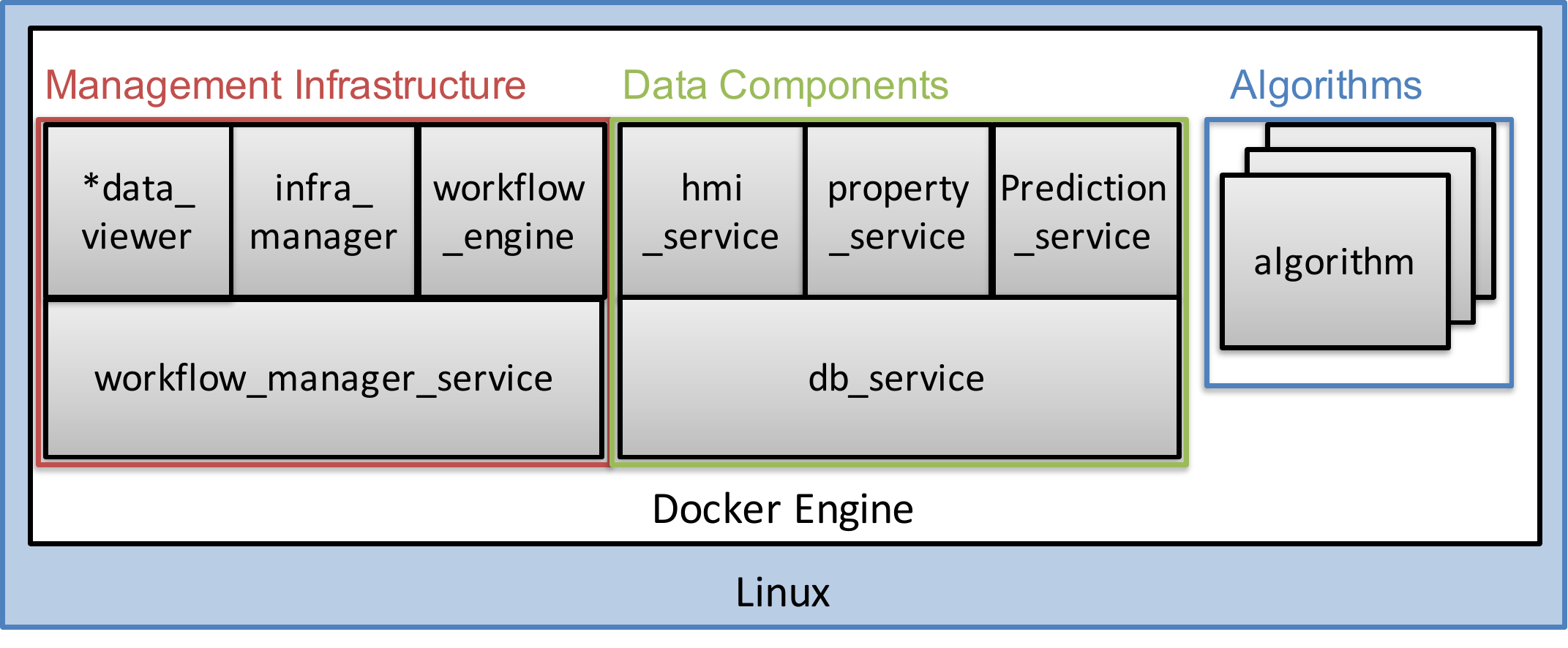}
    \caption{Components of the FLARECAST architecture within the main Docker engine. 
    The different grayscale rectangles correspond to different Docker containers, while adjacent containers typically share a common interface.}
    \label{fig:components}
\end{figure}

All FLARECAST infrastructure components are implemented as Docker containers. 
Docker\footnote{\url{https://www.docker.com}} is an open source software ecosystem (i.e., engine) in which different Docker containers co-exist and function independently. 
A Docker container can be viewed as a light-weight virtual machine that hosts an arbitrarily configured software environment. 
This includes custom programming languages and different library versions per container. 
Each FLARECAST infrastructure component or algorithm is deployed in a dedicated Docker container. 
These containers can be installed on a high performance cluster as well as on a developer's desktop machine, making it possible to deploy FLARECAST in different environments at the same time. 
A schematic of the FLARECAST Docker containers is shown in Figure~\ref{fig:components}. 

\subsubsection{Language independence}
As mentioned already, algorithms may be written in any programming language. 
This allows to re-use already existing libraries for certain tasks. 
Currently tested and supported are Python 2.7 and 3.0, the Interactive Data Language (IDL) 8.5 and R. 
Other languages that can be executed on a Linux system (e.g., Java, C++ and Perl) are not expected to generate major issues, but we must make clear that we have not systematically tested them. 

\subsubsection{Data model}
\label{sec:dm}
The design of the data model is balanced between flexible support for different kinds of data structures and strict schema definitions for better interoperability. 
This can be achieved through a semi-structured data model. 
Its main structures are defined by a traditional table-based, relational data model. 
Individual entries of the table support weakly-typed data types (i.e., types supported by weakly- or loosely-typed programming languages, such as C). 
In this sense, an algorithm can handle data types as simple as numeric values or strings simultaneously with more complex data types, such as arrays, matrices, dictionaries and whole hierarchies of objects. 
For the weakly-typed data types we rely on the built-in JSON data type of PostgreSQL 9.6. 
Algorithms encode their output (e.g., extracted properties) in the JSON data format. 
JSON is supported by most C-like programming languages, including Python, R and IDL. 

% user interfaces?

%\subsubsection {Licensing}
%EU's OpenAIRE policy requires that all source code and data models must be openly accessible to any interested entity, both within the EU and worldwide. 
%Therefore, the software and data licenses chosen by the FLARECAST consortium aim to be as permissive as possible, while neglecting any liability. The majority of the written code is available under a  BSD-2 (\url{https://opensource.org/licenses/BSD-2-Clause}) license. Some of the algorithms, however, depend on third party libraries, that use a less permissive GPL licence (\url{https://www.gnu.org/licenses/gpl-3.0.de.html}) and thus these algorithms need to be released under GPL, as well. All FLARECAST data products are provided under the fully open ODC PPDL (\url{https://www.opendatacommons.org/licenses/pddl/1-0/index.html}) license. 

\subsection{Communication and Dissemination}
\label{s2:comm}
The FLARECAST communication objective ran throughout the project as it had a dedicated, overarching WP (WP7; see the WP list in Section~\ref{s2}). The FLARECAST Consortium aimed to raise awareness in three different directions, namely industry and government, space weather end users, and the general public. 

\subsubsection{Industry / government, scientists and end users}
\label{sec:dissem}
Initial engagement with end users, in terms of both scientists and forecasters, evolved into a User Survey conducted by the Met Office partner in autumn 2016, in advance of the FLARECAST First Stakeholders Workshop held at the Met Office (Exeter, UK) in January 2017. 
In its findings (described in more detail in Appendix~\ref{app2}) about a quarter of respondents were not sure about the accuracy of the forecasts they were using, clearly indicating a need for more education about the forecast methods, as well as about the verification of data, methods, and performance. 
This may also explain why around 60\% of respondents were passive regarding recommending forecast services. 
Further, it was found that timeliness, accuracy and ease of use are the most important factors in a forecast, while scientific detail is the least important information to include. 

The First Stakeholders Workshop was attended by 30 participants in total, including 20 non-FLARECAST attendees from basic solar and space-weather research and operational forecasting, as well as end users and stakeholders from the aviation, defence, marine, satellite, and communications sectors. 
There was discussion of the survey results and an extended discussion on the construction of rough research and development roadmaps for the short and long terms.  Top-level conclusions of the Workshop included the following: 
\begin{itemize}
    \item One forecasting solution is not possible to fit all sectors and users, as different sectors need different forecast windows, latencies and means of verification depending on their generic business models (i.e., false alarms, misses, cost-to-loss ratios, etc$\ldots$). 
    \item FLARECAST forecasts will need an intermediate step before reaching end users, to make the forecasts understandable to them. This step should be taken by operational centres, such as the Met Office Space Weather Operations Centre (MOSWOC). 
    It was deduced that the real end users of FLARECAST are, in fact, the operational forecasters themselves. 
    \item The users stressed the practical aspects of an operational forecast (i.e., reliability and precision) over scientific details. 
    They also found the Workshop to be an excellent educational experience. 
    \item The need for an integrated, Sun-to-Earth, space-weather forecasting system was portrayed as evident and pressing. 
    It will become increasingly pressing in the future.
\end{itemize}
A Second Stakeholders Workshop was held in Ostend (Belgium) in November 2017 (alongside that year's European Space Weather Week) and was oriented more toward scientists, end users and operational space-weather forecasters. 
Among others, it featured participation by the EASA and NASA’s CCMC. 
This second workshop pointed out that FLARECAST is producing and publishing results that are readily available to be used and validated, and the FLARECAST forecasting service will follow shortly. 
However, some convenors agreed that, in some respects, the `fun' for FLARECAST's potential really starts now, since the greater community should think about scientific/operational projects involving the new FLARECAST `software instrument'. 
Finally, in view of the strong contribution from EASA in the discussion, it is increasingly clear that aviation agencies (and the aviation industry, naturally) are becoming more engaged in space-weather impacts and associated forecast services. 
Therefore, the heliophysics community needs to proactively follow up with them to ensure that this opportunity for a meaningful, prolific interaction is not missed. 

In addition, intense communication took place with the scientific community at conferences and during a dedicated FLARECAST Science Workshop organized in June 2017 in Paris by the CNRS partner. 
These actions anchored FLARECAST within the larger field of current and future space-weather research and secured collaboration between interested parties beyond the project duration. 

\subsubsection{Education and public outreach}
A variety of communication formats were used to involve the public in all partner countries: science caf\'{e}s; citizen science events; children’s workshops; interactive exhibits; social media. 
Public engagement activities raised awareness not only about space weather and the need for reliable solar flare forecasts but also, more generally, about the working of collaborative research in Europe. 
We witnessed the relentless interest and fascination of the public on the realization that real efforts are made to predict solar activity, because this can have a very tangible impact on human life and well-being. 
This was not initially clear and we strove to pass the word clearly and compellingly. 

\subsubsection{Press releases and press appearances}

The numerous direct interactions and networking activities between scientists and different stakeholders in the project's framework were aimed at sharing information, advancing mutual understanding, and increasing trust in science (i.e., embedding the project in society at large). 
These efforts came just in time, as two partners (i.e., Switzerland and United Kingdom) faced political uncertainties about their involvement in future European research programmes. 
In one remarkable development{\footnote{\url{https://www.theyworkforyou.com/lords/?id=2016-11-03b.771.2}}}, the project itself was brought up as a paradigm of solid, meaningful collaboration between British and Continental European researchers. 

This visibility was assisted by the Consortium's concerted efforts to exploit opportunities for disseminating the project's achievements and deliberations. 
There were annual press releases in all official languages of the partner countries, while Consortium members were interviewed in numerous occasions by the electronic and printed media based in the partner countries. 
Much attention was, therefore, paid to contextualized press work to make sure that the project was broadly communicated in all participating countries and beyond.

\begin{figure}
    \centering
    \includegraphics[width=\textwidth]{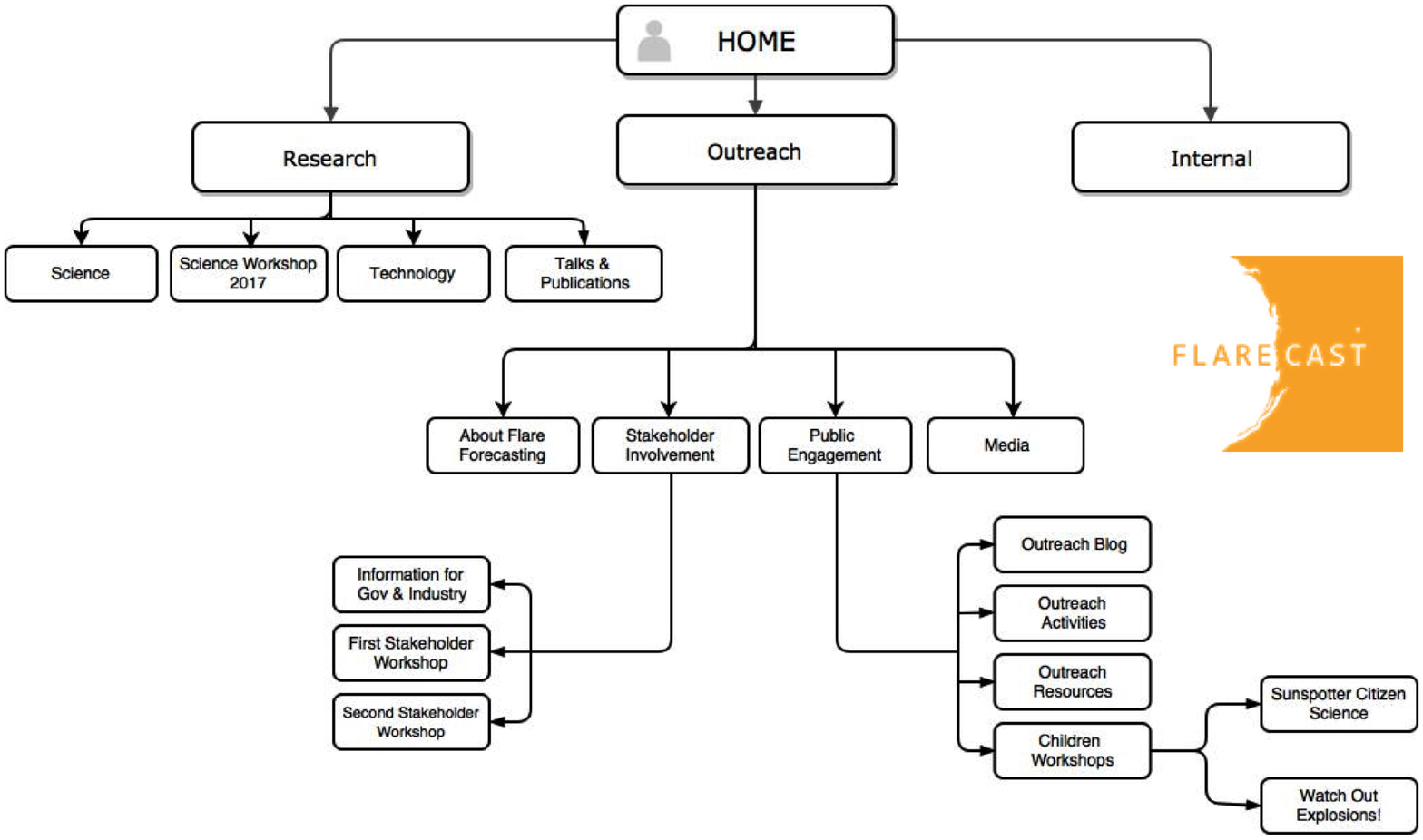}
    \caption{Structure and components of the FLARECAST website (\url{http://flarecast.eu}).}
    \label{fig:website}
\end{figure}

Detailed information on all three communication and dissemination elements of FLARECAST can be found on the project's website, a skeleton diagram of which is provided in Figure~\ref{fig:website}.

\section{The FLARECAST Infrastructure: Handling and monitoring}
\label{s3}
%In this section we provide a schematic description of the mathematical aspects and some comments about the adopted implementation for the methods contained in the current version of the software. In fact, the modular structure of the FLARECAST software allows systematic update of the software status with new machine learning algorithms. It follows that the FLARECAST tool-kit is characterized by a notably dynamic content and new algorithm modules are going to be added to the software during the FLARECAST life and even after the project’s end.

The software developed within FLARECAST is a computational resource enabling one to apply several machine-learning methods to data available in the FLARECAST property database. 
The software is object-oriented, with a set of objects for reading and analyzing data formats (i.e., model learning or prediction), another set for visualizing the results and, finally, another set for storing them in appropriate prediction databases. 
The software has been designed and written in Python, with the pertinent software modules described in the following sections. 

\subsection{Data handling}
Data handling is managed by	a database interface module that can read and collect properties from the FLARECAST property database and write the results of the analysis into other databases. 
The databases involved are:
\begin{enumerate}[leftmargin=*]
    \item   {\bf{Property database:}} It contains the relevant AR information and comprises the following data:
    \begin{enumerate}
        \item  SHARP HMI metadata (i.e., properties);
        \item  data set of SHARP-calculated properties exclusive to FLARECAST; 
        \item  NOAA/SWPC SRS data;
        \item  flare association as extracted from the NOAA/SWPC event list.
    \end{enumerate}
    \item {\bf{Machine learning configuration database:}}	It contains information learned by a given machine learning method from a given training set according to the machine learning algorithm category. 
    We have generated the following machine learning algorithm categories and corresponding parameters to be saved and stored:
    \begin{enumerate}
        \item	Neural Networks $\rightarrow$ architecture, synaptic weights; 
        \item 	Clustering methods $\rightarrow$ clusters’ centers;
        \item	Regression methods $\rightarrow$ predictors’ weights, regularization parameters.
    \end{enumerate}
    \item {\bf{Machine learning result database:}} It contains the prediction results associated to a given AR property set (i.e., the labels predicted by the selected machine learning algorithms). 
    Depending on the machine-learning algorithm utilized, ARs at a given point in time can be labeled using different types of prediction labels. 
    The label types are:
    \begin{enumerate}
        \item	Predicted flare occurrence: binary values, 0 (no-flare) / 1 (yes-flare) values, referring to the flare occurrence at a specific GOES flux level that must be fixed a priori (e.g., flares $\geqslant$M1.0);
        \item   Predicted flaring probability: decimal percentage values between 0 and 1, similar to above;
        \item	Flare intensity: a positive real number;
        \item	Flare intensity and delay: two positive real numbers.
    \end{enumerate}
    The flare intensity number reflects the GOES flare class (forecast and actual), while the delay refers to the interval between the time of the specific flare forecast and the actual flare onset.
    %DSB: Add (at a later time) whether or not these last two results formats are present in our prediction database.
\end{enumerate}

\subsection{Data monitoring}
A consistent and as complete as possible coverage of the SDO/HMI archive at MEDOC is important for FLARECAST to fully exploit the JSOC data, along with the ability to promptly download the latest SDO/HMI NRT SHARPs for as long as they are available.
As a result, both the archive coverage and download delays had to be monitored. Additional monitoring tools have, therefore, been developed to keep track of both the infrastructure workflow and database coverage. 

\subsubsection{Data coverage monitoring}
Monitoring of data coverage is achieved using the scripts in the `coverage' directory of the project's public Git repository{\footnote{\url{https://dev.flarecast.eu/stash/projects/INFRA/repos/hmi-coverage/}}}. 
These scripts are automatically executed daily with their results published at MEDOC{\footnote{\url{http://sdo.ias.u-psud.fr/medoc-hmi-list/}}}. 
This webpage uses color tables to display the percentage of existing SDO/HMI data, for different data series, that are available at JSOC and are mirrored at MEDOC. 
This is done as a function of month and day of month using a color table in which even just one missing observation is visible and encodes in white those cases where no data are available at JSOC. 
Notice that for the SHARP data series a gap may exist either when no eligible SHARPs can be found on the solar disk on a given day or in (relatively rare) cases of issues with the HMI data pipeline. 

\subsubsection{Download delays monitoring}
Monitoring of download delays is achieved using the script in the `delays' directory of the above-mentioned Git repository. 
The observation date and time at which files are available at MEDOC are directly queried from the database, with year and month supplied to the script as external arguments. 

The output is a plot displaying the download delay (compared to the observation [i.e., data acquisition] time) as a function of observing time for the selected month. 
The delay axis spans from 0~hr to 24~hr. 
Missing data (that could not be downloaded from JSOC) are shown as a 0-hr delay in red, while data downloaded with a delay of more than 24~hr are shown as a 24-hr delay in orange.

\subsection{Infrastructure and database coverage monitoring}

%\begin{figure}
%
%    \includegraphics[width=\textwidth]{Figure_10.png}
%    \caption{FLARECAST's viewers for database monitoring: (a) \emph{infra\_viewer}, to monitor and manage the infrastructure (internal); (b) \emph{property\_viewer}, to display the distribution of properties and data gaps (internal); (c) \emph{property\_plotter}, to plot distributions of individual property values (internal); (d) \emph{prediction\_plotter}, to plot predictions of all available algorithms between 2013 and 2018. 
%    Access is available through \url{https://api.flarecast.eu/predictionplotter}.}
%    \label{fig:data_viewers}
%\end{figure}
%
To overview and monitor the workflow process and database coverage, the FLARECAST infrastructure provides several viewers, some of which are intended as internal tools.  These tools can be found in dedicated git repositories in the INFRA project area\footnote{\url{https://dev.flarecast.eu/stash/projects/INFRA}}. The \textbf{\emph{infra\_viewer}} allows FLARECAST developers to manually start new processes and to monitor their performance. The \textbf{\emph{property\_viewer}} and the \textbf{\emph{prediction\_viewer}} allow to visually inspect the distribution and completeness of the data. They help system admins to better assess the quality of the data.

In addition to viewers, the FLARECAST infrastructure provides a collection of API routes to analyse the data on a database level (Appendix~\ref{app1}). 
These scripts iterate through the entire data set and report unexpected gaps in text form. 
Manual analysis is then needed to decide if data need to be reprocessed or the gaps can be explained by missing or corrupted source data (i.e., HMI SHARPs). 

%\begin{figure}
%    \centering
%    \includegraphics[width=0.8 \textwidth]{Figure_11.png}
%     \caption{Input JSON-format configuration file for %\emph{verification\_engine.py}. 
%     This example pertains to the `Hybrid Logit' prediction method used to forecast GOES $\geqslant$C1.0 flares (`abovec') in a 24-hr forecast window with zero latency and a midnight base issuing time. 
%     The property sets to be used are all properties relying on LOS magnetic field without taking flare history into account (`Blos\_all\_nofh') while the verification period corresponds to 1 January 2015 to 31 December 2017. 
%     Uncertainties are calculated via a 90\% bootstrap method implemented with 1,000 realizations.}
%    \label{fig:JSON}
%\end{figure}

\section{Performance Verification Strategy}
\label{s4}
The entire verification software for the FLARECAST project is contained in the \emph{verification\_engine.py} and \emph{verification\_module.py} codes in the project's Stash repository{\footnote{\url{https://dev.flarecast.eu/stash/projects/VER/repos/standard\_verification}}}. 
The starting point for the software is reading from a JSON-format configuration file.  
%(Fig.~\ref{fig:JSON}). 
This initializes the algorithm name for the prediction method, the threshold flare intensity of interest, the base issuing time of the forecasts, the length of the forecast window, the latency of the forecast windows from their issuing times, whether magnetic and/or flare history property sets are used and, finally, the start and end date-times for the period undergoing verification. 
Given an input forecast window length, the software determines which issuing times to be considered for reading prediction `algorithm\_config\_name' entries from the prediction database. 
This is to ensure that temporally edge-to-edge forecasts are considered in the verification, such as a single 24-hr forecast window issued at 00:00~UT each day, or two 12-hr forecast windows issued at 00:00 and 12:00~UT each day, or four 6-hr forecast windows issued at 00:00, 06:00, 12:00, and 18:00~UT each day. 
In all of these examples, the latency (i.e., the time difference between a forecast issuing time and the start of the prediction window) is zero, meaning that forecasts are effective immediately. Nonzero latency can also be implemented, if desired. 

Following this, two separate calls are made to the FLARECAST prediction API. 
The first is a request to the `algoconfig/data' service for the related prediction algorithm configuration settings (or a list of prediction algorithm configuration settings in the case of forecast window durations less than 24~hr). 
The second is a request for the entire set of prediction results through the `prediction/list\_v2' service. 
Having retrieved the full set of forecasts to be verified, minimum and maximum flare class information for that prediction set are extracted from the `algoconfig/data' results. 
This information is needed to convert the associated flare information contained in each prediction result entry into the relevant binary observational truths (i.e., one or more suitable flare did [1], or did not [0], happen within the forecast window) that are then compared to the decimal forecast probability or binary classification issued by the prediction algorithm under consideration.

%SHAUN: This following paragraph is not strictly necessary
%At this stage a loop is performed over each prediction result entry, within which predictions are filtered into separate data structures based on whether they are labeled as probabilities (i.e., decimal values between 0 and 1) or binary classifications (i.e., only 0 or 1). 
%The associated flare information is also included in binary format (i.e., flare did happen [1] or did not happen [0]). 
%Spurious numerical results can occur (i.e., binary values that are close, but not exactly, 0 or 1, or probabilities that are close, but slightly beyond, the range [0,1]). 
%To identify and remove these, both forecast data types are separately screened.  

The next stage of \emph{verification\_engine.py} is the initial construction of the output metadata structure, containing the full set of metadata keys and a placeholder for the `data' key. 
At this point, `forecast\_type' and `source\_predictions' are left blank as the content depends on whether forecasts are probabilities or classifications; distinction is not possible based only on `algorithm\_config\_name' because some prediction methods generate both types. 
Classification forecast-observation pairs are then passed into the \emph{calc\_class\_stats()} function in \emph{verification\_module.py}, while  probability forecast-observation pairs are passed into the equivalent \emph{calc\_prob\_stats()} function. 
Both return structured results saved directly into the `data' key at the metadata level of the output structure. 
Numerous skill scores are then computed for both the probabilistic and the categorical-classification format of forecasts, outlined in more detail below.

\subsection{Probabilistic forecasting}
\label{s:prob}
The computation of probabilistic scores relies on two metrics associated to the $i$-th forecast: the binary observed flaring $o_i \equiv \{0,1\}$, and the forecast probabilty $f_i \in [0, 1])$. 
Using them, the verification module computes the Brier Score, 
\begin{equation}\label{brier-1}
    \mathrm{BS} = \frac{1}{N} \sum_{i=1}^N (o_i - f_i)^2\ ,
\end{equation}
together with its decomposition into the three components of reliability, resolution, and uncertainty \citep[see, e.g.,][]{richardson_2012}. 
From %this, 
the BS, 
one obtains the Brier Skill Score, 
\begin{equation}\label{brier-2}
    \mathrm{BSS} = 1- \frac{\mathrm{BS}}{\mathrm{BS}_{\mathrm{ref}}}\ ,
\end{equation}
where the reference score $\mathrm{BS}_{\mathrm{ref}}$ is given by, 
\begin{equation}\label{brier-3}
    \mathrm{BS}_{\mathrm{ref}} = \frac{1}{N}\sum_{i=1}^N (o_i - \bar{o})^2\ ,
\end{equation}
and $\bar{o}=\sum_{i=1}^N o_i$ is the average of the binary observed flare occurrences over the $N$ forecasts (i.e., the climatology).

Notice that BSS directly combines probabilistic forecasts with binary observations in a paired format. 
Further characterization is achieved through binning of these forecast-observation paired data according to their forecast values. 
This enables construction of the four quantities required to plot reliability diagrams, which are plots comparing the observed frequency of events to the forecast probability
(see, e.g., \citealt{broecker_etal12}, as well as \citealt{murray17} for an application to flare forecasting). 
These quantities are (1) the number of entries in each forecast bin, (2) the average forecast value in each bin, (3) the average observation in each bin, and (4) an estimate of uncertainty in the average observation value. 
The key parameter in the construction of these reliability diagram quantities is the size of the decimal-probability forecast value bins, which is controlled by the `data set'--`probabilities'--`rel\_dia\_stepsize' parameter in the JSON configuration file input into \emph{verification\_engine.py}. 
%(Fig.~\ref{fig:JSON}). 
The reliability diagram data are specifically included in the verification output `data' result structure to prevent the need for the full prediction set to be requested again through the API service at a later time.

In addition to the standard probabilistic metrics, decimal forecast probabilities are converted by a threshold value into categorical-classification forecasts of \emph{no-flare} (i.e., probabilities below the threshold) and \emph{yes-flare} (i.e., probabilities at or above the threshold). 
Varying this threshold yields a series of binary forecast and observation data sets that have the full set of categorical metrics and skill scores calculated for them. 
These threshold-dependent categorical metric data sets are then included in the output probability verification `data' result structure as a nested list of categorical-classification output result dictionaries, with each list entry supplemented by the threshold probability value used to construct that entry. 
These categorical metrics and skill scores are provided below. 

\subsection{Categorical forecasting}
Given a certain range of flare magnitudes, forecast window duration and latency, verification of the simplest possible categorical forecasting scheme (i.e., binary 0 for \emph{no-flare} and 1 for \emph{yes-flare}) relies on the $2 \times 2$ contingency table (aka `confusion matrix') of Table~\ref{tab:cmatrix}. 
This is constructed from the intersection of the binary forecast-observation pairs corresponding to the following four cases: number of forecast windows predicted to flare with flare(s) observed (TP); number of forecast windows predicted to flare with no flare(s) observed (FP); number of forecast windows predicted to not flare with flare(s) observed (FN); number of forecast windows predicted to not flare with no flare(s) observed (TN). 
The first row ($\mathrm{TP}+\mathrm{FP}$) provides the number of flare-predicted forecast windows, while the second row ($\mathrm{FN}+\mathrm{TN}$) corresponds to the number of non-flare-predicted forecast windows. 
The first column ($\mathrm{TP}+\mathrm{FN}$) provides the number of observed flaring forecast windows, while the second column ($\mathrm{FP}+\mathrm{TN}$) corresponds to the observed non-flaring forecast windows. The total number of forecast windows is $N = \mathrm{TP}+\mathrm{FP}+\mathrm{FN}+\mathrm{TN}$. 
Table~\ref{tab:cmatrix} gives rise to a long list of possible forecast metrics and skill scores, with those utilized in FLARECAST outlined in Table~\ref{tab:bmetrics}.

\begin{table}[ht]
   \caption{$2 \times 2$ contingency table for categorical flare forecasting.}
    \centering
    \begin{tabular}{c c | c c}
                        &               & \multicolumn{2}{c}{\textbf{Flaring Observed}}\\
                        &               & \textbf{Yes}          & \textbf{No}\\
    \hline
    \textbf{Flaring}    &  \textbf{Yes} &  True Positive (TP)   &  False Positive (FP)\\
    \textbf{Predicted}  &  \textbf{No}  &  False Negative (FN)  &  True Negative (TN)\\ 
    \hline
    \end{tabular}
    \label{tab:cmatrix}
\end{table}

\begin{table}[ht]
    \caption{Metrics and skill scores used for categorical forecasts in FLARECAST, along with their applicable ranges. All parameters shown rely on the simple $2 \times 2$ contingency table of Table~\ref{tab:cmatrix}.}
    \centering
    \begingroup
    \setlength{\tabcolsep}{5pt} % Default value: 6pt
    \renewcommand{\arraystretch}{1.5} % Default value: 1
    \begin{tabular}{l l c c}
        \textbf{Name}                       &  \textbf{Notation} & \textbf{Formula} & \textbf{Range} \\
        \hline
        Accuracy                            & ACC     & $\frac{\mathrm{TP}+\mathrm{TN}}{N}$                                                         & [0,1]\\
        False alarm ratio                   & FAR     & $\frac{\mathrm{FP}}{\mathrm{TP}+\mathrm{FP}}$                                               & [0,1]\\
        Bias                                & BIAS    & $\frac{\mathrm{TP}+\mathrm{FP}}{\mathrm{TP}+\mathrm{FN}}$                                   & [0,$\infty$]\\
        Threat score                        & TS      & $\frac{\mathrm{TP}}{\mathrm{TP}+\mathrm{FN}+\mathrm{FP}}$                                   & [0,1]\\
        Equitable threat score              & ETS     & $\frac{\mathrm{TP}-R_{\mathrm{ETS}}}{\mathrm{TP}+\mathrm{FN}+\mathrm{FP}-R_{\mathrm{ETS}}}$ & [-$\frac{1}{3}$,1]\\
                                            &         & using $R_{\mathrm{ETS}} = \frac{(\mathrm{TP}+\mathrm{FN})(\mathrm{TP}+\mathrm{FP})}{N}$     & \\
        Probability of detection            & POD     & $\frac{\mathrm{TP}}{\mathrm{TP}+\mathrm{FN}}$                                               & [0,1]\\
        Probability of false detection      & POFD    & $\frac{\mathrm{FP}}{\mathrm{FP}+\mathrm{TN}}$                                               & [0,1]\\
        Odds ratio                          & OR      & $\frac{\mathrm{TP} \cdot \mathrm{TN}}{\mathrm{FN} \cdot \mathrm{FP}}$                       & [0,$\infty$]\\
        Odds ratio skill score              & ORSS    & $\frac{(\mathrm{TP} \cdot \mathrm{TN})-(\mathrm{FN} \cdot \mathrm{FP})}{(\mathrm{TP} \cdot \mathrm{TN}) + (\mathrm{FN} \cdot \mathrm{FP})}$ & [-1,1]\\
        Heidke skill score                  & HSS     & $\frac{\mathrm{TP} + \mathrm{TN} - R_{\mathrm{HSS}}}{N - R_{\mathrm{HSS}}}$                 & [-1,1]\\
                                            &         & using $R_{\mathrm{HSS}} = \frac{(\mathrm{TP}+\mathrm{FN})(\mathrm{TP}+\mathrm{FP})+(\mathrm{TN}+\mathrm{FN})(\mathrm{TN}+\mathrm{FP})}{N}$ & \\
        True skill statistic                & TSS     & $\mathrm{POD} - \mathrm{POFD}$                                                                                & [-1,1]\\
        Symmetric extremal dependence index & SEDI    & $\frac{\log(\mathrm{POFD})-\log(\mathrm{POD})-\log(1-\mathrm{POFD})+\log(1-\mathrm{POD})}{\log(\mathrm{POFD})+\log(\mathrm{POD})+\log(1-\mathrm{POFD})+\log(1-\mathrm{POD})}$ & [-1,1]\\
        Appleman's discriminant             & AD      & $\frac{\mathrm{TN}-\mathrm{FN}}{\mathrm{FP}+\mathrm{TN}} \quad \mathrm{if} \quad (\mathrm{TP}+\mathrm{FN}) > (\mathrm{FP}+\mathrm{TN})$ & [-$\frac{\mathrm{FN}}{\mathrm{FP}}$,1]\\
                                            &         & $\frac{\mathrm{TP}-\mathrm{FP}}{\mathrm{FN}+\mathrm{TP}} \quad \mathrm{if} \quad (\mathrm{TP}+\mathrm{FN}) < (\mathrm{FP}+\mathrm{TN})$ & [-$\frac{\mathrm{FP}}{\mathrm{FN}}$,1]\\
\hline
    \end{tabular}
\endgroup\
\label{tab:bmetrics}
\end{table}

Another important element of forecast verification is the use of the POD and POFD (see Table~\ref{tab:bmetrics}) as ordinate and abscissa, respectively, of the ROC plot (see, e.g., \citet{broecker_etal12}, as well as \citet{sharpe_murray17} for an application specific to flare forecasting). 
For thresholded probabilistic forecasts \citep{bloomfield12} the ROC plot contains the (POD, POFD) pairs for each probability threshold. 
In a manner similar to the reliability diagram of Section~\ref{s:prob}, readily creating ROC plots avoids unnecessary reprocessing of large quantities of forecast data just for visualization purposes. 
ROC plots directly provide the AUC metric that is determined by the integration of the POD values over their spacing in POFD. 

\subsection{Verification metric uncertainty calculation}
\label{s:uncert}
In standard operation, the verification process calculates a single value of each metric and skill score that represents the performance achieved across all forecasts within the selected verification time range.
FLARECAST also includes an optional assessment of uncertainty for these single-valued metrics and skill scores, controlled through the `data set'--`uncertainties' parameters in the JSON verification configuration file. 
%(Fig.~\ref{fig:JSON}). 
%This includes a Boolean `calculate' switch that is checked by \emph{verification\_engine.py} in order to either perform (`True’) or not perform (`False’) the uncertainty assessment. 
The choice of resampling method for the calculation of uncertainties is controlled by the string-format `sampling\_scheme' parameter. 
In the current implementation, three resampling schemes can be utilized, namely bootstrap and jackknife \citep{1982jbor.book.....E} or random sampling without replacement (i.e., a sub-sampling method). 
The size of each partial sample is set via `sample\_percent', namely the decimal percentage of the total number of forecast-observation pairs in the verification time range, $N$. 
Finally, the number of times that resampling is performed is set by the `realizations' parameter. 
It is worth noting that the jackknife method overrides these to self-define the sample size as $N-1$ (one less than the number of forecast-observation pairs) and the number of realizations as $N$, because this is the number of times one %data point 
forecast-observation pair can be dropped. 
In addition, the standard bootstrap method uses $N$ samples (i.e., `sample\_percent' = 1), with the freedom to choose any number of realizations. In our infrastructure it is also possible to run the FLARECAST verification engine with the bootstrap resampling with replacement but using `sample\_percent' $<1$, although this should not be treated (or referred to) as a standard bootstrap implementation, except in the case `sample\_percent' = 1.  

The assessment of metric uncertainties is optionally invoked after the categorical-classification or probability forecast metrics have been written into the output verification data structure. 
In the \emph{calc\_ss\_uncert()} uncertainty assessment function, indices for the forecast-observation pair arrays are resampled to create 2D index arrays of the form [`realizations', `sample\_size']. 
For a bootstrap, indices in the range $[0, (N-1)]$ are uniformly randomly drawn for a total of `realizations' $\times$ `sample\_size' times %by \emph{numpy.random.choice()} 
 to directly populate the 2D index array. 
For a jackknife, each index is dropped to generate a 2D $[N, (N-1)]$ index array. 
Finally, for a subsample, in each realization the monotonic array of indices $0, 1, 2, ..., (N-1)$ are randomly shuffled %by \emph{numpy.random.permutation()} 
 and the first `sample\_size' number of indices are extracted and stored in the 2D [`realizations', `sample\_size'] index array. 
After the 2D index permutation sets are generated, the full suite of verification metrics and skill scores are calculated over the `sample\_size' entries for each realization. 
These are achieved using \emph{calc\_class\_stats()} and \emph{calc\_prob\_stats()} once again to ensure consistency of the metric and skill score calculations with those reported as primary results in the verification data structure. 
To simplify the uncertainty assessment portion of the structure, any key names containing nested structures are removed (i.e., the contingency table in the classification data structure and the probability-thresholded reliability diagram in the probability data structure). 
Following this, averages and standard deviations are calculated over the number of realizations for each separate metric and skill score. 
The resulting dictionaries of average and standard deviation values are separately saved into the `unc\_avg' and `unc\_std' keys, respectively, in the output verification data structure. 

We note for completeness that the above implementation of metric uncertainties is used because it is deemed as the most time and computationally effective. Alternative  implementations can be pursued, but these are reserved for future FLARECAST infrastructure upgrades.

%\subsection{Predictive models}
%\label{s4:pred}

%A machine learning algorithm launcher module, which can execute the training phase of the machine learning algorithm in use applied to a given labeled AR feature set and can execute the prediction given an un-labeled AR feature set. The machine learning algorithms currently implemented in the software are:
%\begin{enumerate}
%\item Clustering techniques: K-Means %\citep{macqueen67}, Fuzzy C-Means %\citep{ruspini69,bezdek84}, Possibilistic C-Means %\citep{pcm1993,pcm1996}.
%\item Classification via regression methods: Hybrid Lasso \citep{benvenuto2018hybrid}, Logit (see e.g. \cite{friedman2001elements}), Support Vector Machines (see e.g. \cite{scholkopf2001learning}), Multi-task Learning for count data \citep{guastavino2019consistent}.
%\item Neural Network: Multi Layer Perceptrons (see e.g. \cite{bishop2006pattern}).
%\item Random Forest: \cite{breiman2001random}. Early applications of the random forest algorithm in solar flare forecasting are \citep{Barnes:16,Liu:17,Florios:18}. {\color{red} Kostas Florios edited here.}
%\end{enumerate}

\section{The FLARECAST Science Results} 
\label{sec:WP6}
FLARECAST's scientific outcome relied on the following four pillars: an investigation of new, or a revisit of potentially interesting, flare predictors (Section~\ref{sec:NewPredictor}); the application of machine-learning methods to flare forecasting, also enabling feature (i.e., predictor) ranking in terms of predictor importance (Section~\ref{sec:ML-features}); exploration of, first, eruptivity studies with synthetic, 3D MHD models (Section~\ref{sec:synthetic}) and, second, the flare-CME connection (Section~\ref{sec:FlareCME}). This last component shows that, importantly, the FLARECAST AR property database can be meaningfully coupled with CME catalogs such as those of the EU HELCATS project\footnote{{\url{https://www.helcats-fp7.eu/}}} or NASA's DONKI\footnote{{\url{https://swc.gsfc.nasa.gov/main/donki}}} to extend flare forecasting to eruptive flare (i.e., CME onset) forecasting with expectation values of CME properties. 
The complete list of FLARECAST-related peer-reviewed articles can be found in Appendix~\ref{app3}. 

\subsection{New and revisited flare predictors} 
\label{sec:NewPredictor}
%\textcolor{magenta}{Sung-Hong: I briefly described some of flare predictors which have been newly developed and implemented in FLARECAST, in particular with respect to shear flows (I can also mention some other predictors, e.g., parameters associated with magnetic helicity flux, Poynting flux, decay index and null points if needed). Ioannis, Costis and Jordan, please add more predictors that you developed/tested.}

%EP  Here we briefly describe flare predictors which have been developed in FLARECAST and examined to see whether they posses a high potential as improved flare predictors. Among them, we first introduce parameters derived from a map of plasma shear flows for a given AR. Strong shear flows are often observed along magnetic polarity inversion lines (MPILs) on the photospheric surface of flaring ARs. An algorithm was developed to quantitatively characterize photospheric shear flows in ARs \citep[refer to ][for the details]{park18} with three parameters of the mean ($<$\,$S$\,$>$), maximum ($S_\mathrm{max}$) and integral ($S_\mathrm{sum}$) shear-flow speeds along strong-gradient, strong-field MPIL segments. The developed algorithm was then applied to a large data set of $\sim$2,500 co-aligned pairs of AR vector magnetograms with 12-min separation over the period 2012--2016. It was found that ARs with more widespread and/or stronger shear flows tend to not only be more flare-productive in the next 24~hr, but also produce major flares sooner.

In spite of possessing one of the most comprehensive ensembles of flare predictors, FLARECAST also opted to identify additional properties that could assist in flare forecasting. In this respect, subsets of the FLARECAST property (including flare association) database were used and linked to the flaring history of ARs. Promising quantities involved photospheric shear flows \citep{park18}, non-neutralized electric currents \citep{kontogiannis17}, magnetic gradients \citep{kontogiannis18}, and the differential emission measure calculated above ARs \citep{Gontikakis_etal20}. 
In addition, FLARECAST has taken advantage of the results of \citet{Guerra2018}, advocating for the joint use of line-of-sight and solar-surface-vertical versions of predictors in machine learning methods, as well as of the results of \citet{mccloskey_etal16}, revisiting sunspot class associations with flaring rates and confirming the increased likelihood of flaring in case flux emergence enhances the photospheric compactness of ARs.

%The first approach carried within the FLARECAST project to determine interesting flare predictors consisted in using observational data: a subset of the FLARECAST database was used in order to investigate the possible link between innovative quantities and the flaring history of the sampled ARs. The new quantities tested are related to photospheric shear flows \citep{park18}, electric currents \citep{kontogiannis17} as well as magnetic gradients \citep{kontogiannis18}.

Strong shear flows are observed along intense magnetic PILs in the AR photosphere. 
PILs are arguably the single most telltale feature of enhanced flare productivity. 
Properties derived from a map of plasma shear flows were thus analysed by \citet{park18}. 
In particular, an algorithm was developed to quantify photospheric shear flows in ARs in terms of three properties, treated as predictors in this framework: the mean ($<$\,$S$\,$>$), maximum ($S_\mathrm{max}$) and integral ($S_\mathrm{sum}$) shear-flow speeds along strong-gradient, strong-field PIL segments. 
This algorithm was then applied to a large data set of $\sim2,500$ co-aligned pairs of AR vector magnetograms with 12-min separation over the period 2012--2016. 
\citet{park18} found that ARs with more widespread and/or stronger shear flows tend to not only be more flare-productive over the next 24~hr but also produce major flares sooner. 
This finding paves the way for future conditional predictions as well as forecasts better integrating the flare history of ARs. 
The importance of flare history in predictions of future flaring has already been identified by \citet{2012ApJ...757...32F}, as well as by comprehensive statistical studies \citep{leka_etal19a, leka_etal19b, park_etal20}. As further discussed in \citet{park18}, \citet{Welsch2009} found that shear-flow properties exhibited a weaker, albeit still positive, correlation with the flare peak flux than Schrijver's R. However, \citet{park18} revisited the problem using almost as large a sample of co-aligned magnetograms as \citet{Welsch2009}, but at much higher quality (i.e., spatial resolution, cadence) and with a more detailed sample of NOAA-numbered ARs (i.e., in terms of observed times/locations, flaring activities and photospheric magnetic properties). \citet{park18} further focused on flaring activity over the next 24 hours, as well as on the waiting time between major flares as a function of the shear-flow properties. We believe that the new results obtained with this improved magnetogram dataset and systematic study warrant additional investigation.

\begin{figure}[t]
    \centering
    \includegraphics[width=\textwidth]{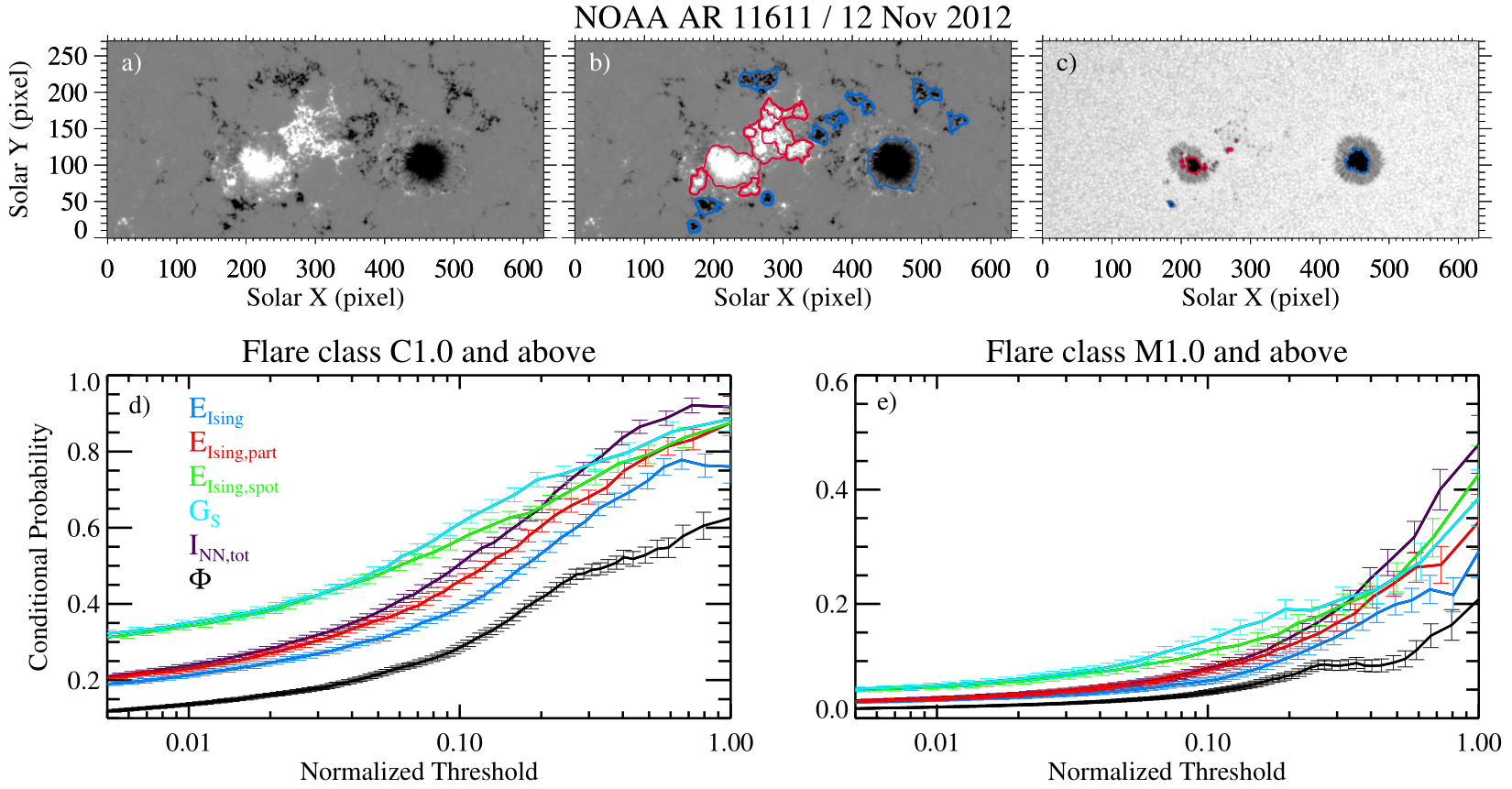}
    \caption{Example of the development and tests of new morphological flare predictors (adapted from \citet{kontogiannis18}). 
    (a) Input LOS magnetogram of NOAA 11611. 
    (b) The magnetogram is partitioned into positive- (red contours) and negative- (blue contours) polarity flux patches that are used to derive the Ising energy $E_{\mathrm{Ising,part}}$ of the AR. 
    These partitions, along with the vector magnetogram of the AR, are further used to calculate the total unsigned non-neutralized currents $I_{\mathrm{NN,tot}}$. 
    (c) Positive- (red) and negative- (blue) polarity umbrae identified in maps of continuum intensity are used to calculate the Ising energy of the umbrae $E_{\mathrm{Ising,spot}}$ and the sum of the horizontal magnetic gradient $G_{s}$. 
    (d)-(e) Conditional probabilities for GOES $\geqslant$C1.0 and $\geqslant$M1.0, respectively,  calculated for a set of successive thresholds in $E_{\mathrm{Ising}}$, $E_{\mathrm{Ising,part}}$, $E_{\mathrm{Ising,spot}}$, $G_{s}$ and $I_{\mathrm{NN,tot}}$. 
    The respective probabilities inferred for the total unsigned magnetic flux $\Phi$ are also shown for reference.}
    \label{fig:new_pred}
\end{figure}

The predictive potential of properties related to non-neutralized electric currents in ARs was also investigated. 
Intense, shear-ridden PILs are exclusive areas of non-neutralized currents that imply the injection of net currents into the corona whose associated non-potential (i.e., free) magnetic energy could be instrumental for flares and eruptions \citep{Torok14,Dalmasse15}. 
\citet{geo_titov_mikic12} had already proposed a method to calculate non-neutralized currents accurately, along with their applicable uncertainties.
%minimizing the effects of measurement errors and numerical uncertainties. 
This method is slightly different than the one of \citet{leka_barnes03a} who added the absolute values of the total current from each magnetic polarity in that, first, it algebraically adds the total currents from both polarities to obtain the net current and, second, it performs a robust flux partitioning of a given magnetogram to identify and distinguish between magnetic polarities. The integral form of Ampere's law for the inference of net currents was also applied by \citet{falconer_etal02}; however, no specific discussion on the precise choice of contours was made. Relying on the method of \cite{geo_titov_mikic12}, \citet{kontogiannis17} implemented the computation of the non-neutralized electric currents within the FLARECAST property database. 
%aiming to produce relevant flaring predictors suitable for automated forecasting systems. 
Calculations for a sample of AR time series showed with statistical significance that the systematic increase in the amount of non-neutralized currents concurred with the development of strong PILs and signaled phases of intense flaring. 
Further application to a representative sample of Solar Cycle 24 ARs showed that the total unsigned non-neutralized current of an AR is  promising in distinguishing between flaring and non-flaring AR populations within a 24-hour time window \citep{kontogiannis17}. 'Promise' in that work implied a significantly better (i.e., beyond applicable uncertainties) ability to distinguish between flaring and non-flaring active regions than the (often used as baseline) unsigned magnetic flux -- see Figure \ref{fig:new_pred} and discussions in \citet{barnes_leka08,georgoulis12,Georgoulis_2013}.

Additionally, FLARECAST developed new algorithms to reproduce some recently proposed predictors not yet implemented in forecasting schemes, relying on the complexity of the spatial distribution of the photospheric magnetic flux. These were the sum of the horizontal magnetic gradient \citep[c.f.][]{korsos14,korsos15,korsos16} and the Ising energy \citep{ahmed10}. 
The sum of the horizontal magnetic gradient is calculated by combining magnetograms and photospheric continuum images. 
The latter are used to locate the umbrae of sunspot groups while the magnetograms are used to define the mean magnetic field strength in the umbrae. 
These values and the separation distances between opposite-polarity umbrae are used to calculate the sum of the magnetic field gradients between all possible opposite-polarity umbrae pairs. 
Furthermore, while the original formulation of Ising energy considers pairings between opposite polarities on a pixel-by-pixel level, FLARECAST examined that form as well as the Ising energy of opposite-polarity umbrae and topologically inferred magnetic partitions. 

\citet{kontogiannis18} demonstrated that magnetic-gradient- and Ising-energy-based predictors are also worth including in automated flare forecasting schemes. 
In terms of Bayesian conditional flaring probabilities (Fig.~\ref{fig:new_pred}), the sum of the horizontal magnetic gradient appeared to be the most efficient property, followed by the Ising energy of the umbrae pairs. Efficiency here implies again an ability to better distinguish between flaring and non-flaring active region populations, not necessarily in terms of predictive potential. The latter, of course, is to be tested in more strict terms within machine learning prediction methods employing multiple predictors and being able to rank them. Although many of these and other properties relate to PILs, \citet{kontogiannis18} showed that each of these properties exhibits specific dependence on the position of ARs on the solar disk and encompasses different information. These differences can justify the different efficiency of these flare predictors and their potential inclusion in multi-parameter machine learning-based forecasting schemes.

%\textcolor{magenta}{Costis: here's a first draft to describe the calculation of DEM timeseries, computed with AIA images and how we can predict solar flares.}

\citet{Gontikakis_etal20}, in a forthcoming study, introduced a new, `orthogonal' view on flare prediction, analysing EUV image time series of ARs. 
More specifically, the temporal evolution of the DEM, calculated over ARs, was  studied as a possible flare predictor. 
The DEM shows the distribution of the plasma EM as a function of temperature, $T$, and is derived from optically thin spectral lines. 
It is defined as $\mathrm{DEM}(T) =  n_e^2 dl/dT$ and involves the derivative of the line-of-sight $l$ over the plasma temperature, with $n_e$ being the electron number density. 
\citet{Syntelis_etal16} found that the DEM calculated over an intensely eruptive AR increased substantially a few (i.e., $\sim5$) hours before one of the eruptions, with the other eruption following shortly after. 
This was interpreted as a possible pre-flare manifestation \citep[see, e.g.,][]{Fletcher_etal11}. 
To achieve statistical significance in a DEM-based pre-flare activity study, \citet{Gontikakis_etal20} used DEM inferences by the GAIA-DEM archive\footnote{The GAIA-DEM is a MEDOC database that archives full-disk maps of DEM parameters, derived from the 6 EUV channels of the SDO/AIA telescope.}. 
In GAIA-DEM, the DEM is assumed to be a Gaussian function of $\log(T)$ \citep{2012ApJS..203...25G} and the archive provides images of three Gaussian parameters, along with $\chi^2$. These parameters are the EM (i.e., the integral of the DEM), the temperature $T_{\mathrm{max}}$ where the DEM is maximized, and the Gaussian width of the DEM. \citet{Gontikakis_etal20} used 6-hr long time series of each DEM parameter.
The DEM parameters were computed from solar structures extracted from 9,000 HMI HARPs. 
Positive derivatives of the time series, indicating that AR plasma heating is at work, were treated as indicators of imminent flares. 
Analyzing the DEM time series appeared to give more significant conditional flare probabilities than the reference unsigned magnetic flux (i.e., its value at the end of the time series) for GOES $\geqslant$M1.0 flares, but it was less successful for GOES $\geqslant$C1.0 flares. 
These may be grounds to advocate that, provided that the significant uncertainties in the DEM calculation can be constrained, DEM time series may complement or even enhance the short-term predictive ability of photospheric properties, at least for major flares. 

%EM and $T_{\rm max}$ were kept for this study as their calculations is less affected by noise.

%The probability for flare occurrence up to 24~hr after the end of the timeseries, under the condition that the derivative of the timeseries is positive, was computed to check the robustness of the prediction method. The conditional probabilities \citep{2012SoPh..276..161G}, computed using  the DEM timeseries were compared with conditional probabilities derived from the unsigned magnetic flux ($\Phi$)  from the same FOVs, observed at the end time of each timeseries. 
%For the case of major flares (>M1), taking place in a time window of 2~hr, 6~hr and 12~hr, the probabilities computed using the EM and $T_{\rm max}$ timeseries were higher than the probabilities computed using $\Phi$. When all flares are taken in the calculation of probabilities, (>C1) the DEM timeseries are less successful. The EM conditional probabilities are higher for the 2-hr and 6-hr time windows from the $\Phi$ conditional probabilities. For the 24-hr time windows the DEM conditional probabilities are lower than the respective $\Phi$ ones. 
%The finding that the conditional probabilities, derived from the DEM parameters are higher than the probabilities computed using $\Phi$ are indicating that the DEM parameters can be considered as a possible flare predictor for time windows smaller than 24~hr.

\subsection{Flare prediction and feature ranking}
\label{sec:ML-features}
The FLARECAST machine-learning prediction component draws from and relies on the plethora of predictors extracted from NRT HMI SHARP magnetograms  populating the project's property database. 
This central objective has been realized by \citet{benvenuto2018hybrid}, \citet{Florios:18}, \citet{pietal19} and \citet{2019ApJ...883..150C}, while the overall process has been described and discussed by \citet{massone_etal18}.

\citet{benvenuto2018hybrid} tested an array of supervised and hybrid (i.e., comprising both supervised and unsupervised elements; see Section~\ref{s2:ml}) machine-learning methods on historical NOAA/SWPC sunspot classifications between August 1996 and December 2010, training the methods on similar data between December 1988 and June 1996. 
They relied on flaring rates associated to certain sunspot classes, similar to \citet{mccloskey_etal16}, but using machine learning for prediction. 
The main result was that hybrid methods tend to outperform strictly supervised ones and approach the performance of clustering methods. 
\citet{benvenuto2018hybrid} used a variety of metrics and skill scores present in Table~\ref{tab:bmetrics} and further determined that a reliable feature ranking by means of their prediction value is possible. 

\citet{Florios:18} used three different supervised methods (i.e., MLPs, SVMs and RFs; see Table~\ref{tab:ML-methods}) on a subset of the FLARECAST property database over a five-year interval covering 2012--2016. 
They were able to correlate between the different methods and concluded that RFs \citep{breiman2001random} could be the method of choice in a routine, operational flare forecasting scheme. 
Feature ranking was performed by means of different scores and indices and it was found that peak performance could be achieved by using $10-12$ predictors in total. 
Both purely probabilistic and probability-thresholded categorical performance verification was performed, with optimal probability thresholds determined. 
The overall performance practically matched, and in some cases exceeded, that of \citet{bobra_couvidat15} who used a SVM and performed the first flare forecasting work using HMI SHARPs. 
%Several metrics and skill scores were used from Table~\ref{tab:bmetrics} but emphasis was placed on the TSS and HSS metrics. Top TSS and HSS values achieved for the forecasting of GOES flare classes $>$ M1 were of order $\sim 0.75$ and $\sim 0.50$, respectively. 

The study of \citet{2019ApJ...883..150C} performed a data- and property-intensive flare prediction investigation. It utilized a database of 14,931 point-in-time property vectors, each comprising up to 171 predictors within the interval September 2012 to April 2016. 
The main objective was to determine whether, and to what extent, this abundance of predictors is essential or redundant in flare prediction. 
The study utilized two machine-learning methods, namely hybrid LASSO and RFs, and opted to construct four training and four testing sets, each corresponding to a 24-hr forecast window from a specific issuing time (i.e., 00:00, 06:00, 12:00, and 18:00~UT each day). 
Particular attention was paid to the complete separation between training and testing sets, that were not only non-time-overlapping (i.e., results from each issuing time were not combined) but they used different ARs (i.e., HARP numbers, in implementation) for training and testing. 
As such, another objective was to find the impact of different intervals of the \emph{same} ARs being used in both training and testing. 
The main findings of \citet{2019ApJ...883..150C} were the following: 
\begin{itemize}
    \item Properties with the best predictor ranking have the smallest variance in ranking, which implies that their impact on prediction is consistently high. However, the order of best-ranking predictors varies for different prediction methods.
    \item Particularly for prediction of GOES $\geqslant$C1.0 flares, the best-ranking predictors are common across all issuing times studied. This is less of a case for the prediction of major flares of GOES $\geqslant$M1.0 flares.  
    \item Only a small number of predictors (different for each method, however, most notably for major flares) suffices to make the prediction method achieve its top performance. Hence, the vast majority of predictors are redundant in such approaches and, in fact, using all of them can even be detrimental to performance. The redundancy of predictors was also noted in \citet{barnes_etal16} as possibly due to intrinsic, but not necessarily known ab initio, correlations between different predictor values. This effect may also be enhanced for events of increased rarity and hence poorer statistics, i.e., when forecasting increasingly higher flare classes.
    \item The stricter approach of separating ARs between training and testing sets leads to notably lower skill score values. Relaxing the robustness of training and testing by allowing `information leaking', not necessarily in terms of temporal overlapping but in terms of different intervals in the same ARs, readily increases these values. Improved performance in this case, however, is due to memorization, rather than learning, on behalf of the machine learning method used. Lower skill scores for unbiased training and testing, as also found by \citet{pietal19}, reflect the intrinsic, uncompromising stochasticity of the flaring phenomenon. 
\end{itemize}

We note at this point that the relatively small number of best predictors found by \citet{Florios:18} and \citet{2019ApJ...883..150C} essentially aligns with earlier results by \citet{ahmed13,bobra_couvidat15} and \cite{barnes_etal17}. The latter study also found that a set of three or four predictors was sufficient to guarantee best performance for two classifiers, NPDA and RF. That study addressed prediction of both filament eruptions and flares and found significant overlapping of about half of best predictors between the two types of predictions. In the case of \citet{2019ApJ...883..150C}, however, such predictor patterns were found only for the prediction of GOES $\geqslant$C1.0 flares within different classifiers, while mostly different sets for different classifiers were found in the prediction of major flares (i.e., GOES $\geqslant$M1.0).

\begin{figure}
   \centering
    \includegraphics[width=\textwidth]{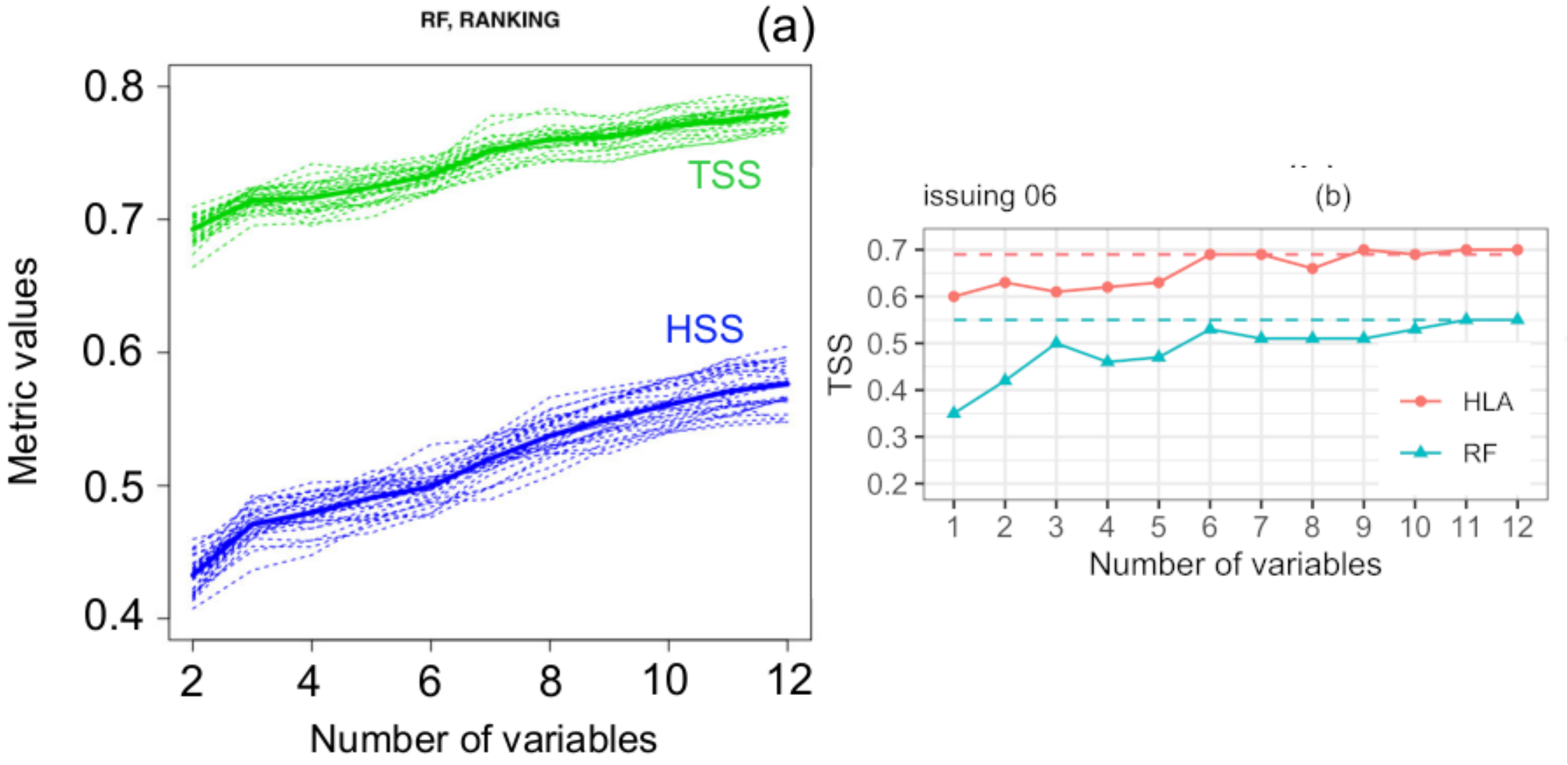}
    \caption{Example prediction runs adapted from (a) \citet{Florios:18} and (b) \citet{2019ApJ...883..150C} for the same basic prediction settings, namely a 24-hr forecast window for GOES $\geqslant$M1.0 flares. 
    Both TSS and HSS values are shown in \emph{a} using RFs only, while \emph{b} shows only TSS values for two different methods, namely hybrid LASSO (HLA) and RF. 
    The abscissas in both plots correspond to the number of predictors used.}
    \label{fig:ml_forecasting}
\end{figure}
Both \citet{Florios:18} and \citet{2019ApJ...883..150C} used various metrics and skill scores from Table~\ref{tab:bmetrics} but focused on HSS and TSS. 
Example runs by both studies and their associated HSS and TSS values are shown in Figure~\ref{fig:ml_forecasting}. This leads to the conclusion that, 
for both studies, the metric values increase while adding more predictors, although in both cases the amplitude of the increase and the values achieved depend on the adopted skill score. Furthermore, in \citet{Florios:18} the increase seems to persist after 12 predictors while in \citet{2019ApJ...883..150C}  saturation is reached with just six predictors (although when RF is implemented on three predictors, it leads to a TSS value very close to the one obtained with six predictors). The difference between the two studies (and subsequent panels in Figure \ref{fig:ml_forecasting}) is probably due to the different way the training set is populated, the latter experiment not mixing up active regions  between the training and test sets.
%In both \mg{machine learning models of Figure \ref{fig:ml_forecasting}b} we see the performance practically peaking after $\sim10$ predictors are used. 
We notice, moreover, that the TSS value for \citet{2019ApJ...883..150C} is lower than that of \citet{Florios:18} for both applied methods, despite the fact that the predictors used in \citet{Florios:18} are a subset of those used in \citet{2019ApJ...883..150C}. 
This is again a result of the more stringently separated training and testing procedure followed by \citet{2019ApJ...883..150C}. 
In fact, the prediction method itself may often be less important than the robustness of training and testing practices. 
By roughly reproducing the training and testing of \citet{bobra_couvidat15} and \citet{Florios:18}, \citet{2019ApJ...883..150C} were able to practically match the performances of these studies for different machine learning methods. 
As a result, the `uncompromised' stochasticity of flare occurrence, as coined by \citet{2019ApJ...883..150C}, could not be curbed even with one of the largest-dimensionality parameter spaces.

Given the above, the FLARECAST performance verification infrastructure (Section~\ref{s4}) is in place and allows the evaluation of performance of any future effort than can rely on either the existing or enhanced property database (Section~\ref{data}) and the prediction database (Section~\ref{s2:ml}). Obviously, the quest continues. 

At this point it is important to clarify that,  although FLARECAST valued highly the use of predictor time series for flare forecasting, this important part of the analysis did not have the time to materialize during the project's nominal period of execution. 
The only exception is the forthcoming work of \citet{Gontikakis_etal20} that exploits DEM time series to uncover flare predictors or precursors. Time series, in terms of flare history and / or predictor evolution, have been found useful in flare prediction \citep{leka_etal18,leka_etal19b,park_etal20} on top of point-in-time forecasting. While this avenue is yet to be investigated with the FLARECAST databases, these data clearly enable one to pursue this objective in the future.

\subsection{Studies using synthetic data}
\label{sec:synthetic}
%\subsection{Diagnostics of eruptivity in solar active regions -- Etienne}

While observational data are the ultimate reference frame to determine the flare-predictive ability of a given quantity, these data have critical limitations toward a complete and pertinent description of solar ARs. For example, while it is agreed that the 3D coronal distribution of the AR magnetic field plays a key role in the triggering and development of instabilities leading to flares and eruptions in general, only the 2D photospheric field vector can be routinely inverted from Stokes images by means of the Zeeman effect.
Thanks to tremendous developments in computational solar physics, numerical experiments of flare and eruption triggering are becoming increasingly  common and realistic \citep[see, e.g., the review of][]{Green18}. 
These experiments have now made it possible to explore new flare predictors. 
\citet{Kusano12} have carried out numerous 3D MHD simulations in which the initial set-up was parametrically modified to accommodate eruptive and non-eruptive configurations. 
Distinguishing between flaring and non-flaring configurations is essential in investigating the efficiency of potential predictors. 

This exploratory component of FLARECAST was facilitated by the 3D MHD simulations of \citet{Zuccarello18}, who extended upon  \citet{Kusano12}. 
These simulations started with a dipole magnetic configuration, emulating the two polarities of an AR, and were driven by specific, carefully modeled and controllable photospheric velocity flows. 
Four types of converging flows were tested (all eventually leading to eruptions) against a control, non-eruptive case. 
Depending on these driving flows and their experimental control, eruptions took place at different times that, however, could be determined precisely for
%Given the experimental control of driving flows, it was made possible to determine precisely at which point the system became unstable and erupted in 
each simulation. 
Given the full 3D data sets, different scalar volumetric quantities were calculated, among which were the potential and non-potential (i.e., free) magnetic energy budgets and the relative magnetic helicity $H$, that was further decomposed into non-potential and volume-threading helicity terms \citep[c.f.][]{Berger03,Moraitis14,Linan18}. 
At the onset of eruptions, all parameter values showed a significant dispersion, indicating the lack of a practical instability threshold. 
Interestingly, no particular free-energy threshold could be identified for flares and eruptions in general, despite the key role of the magnetic free energy in eruptions. 
However, the ratio of a specific term $H_j$ (coined the `non-potential' helicity) to the total relative helicity behaved differently; eruptions seemed to always take place for the same value of the ratio $H_j/H$. Therefore, this ratio may represent a critical threshold at which a modeled AR becomes unstable. 

The importance of the ratio $H_j/H$ was further highlighted by \citet{Pariat17} who analysed parametric 3D MHD simulations of the formation of solar-like ARs. 
These simulations were originally presented by \citet{Leake13,Leake14} and modeled the self-consistent formation of ARs via the emergence of a twisted magnetic flux tube from the upper convection zone into the solar atmosphere. 
Flux emergence led to the formation of a magnetic configuration similar to standard bipolar ARs. 
The parametrisation of the initial conditions enabled the simulation of both quiescent and eruptive configurations. 
In \citet{Pariat17}, the full 3D MHD simulated coronal magnetic field for different cases was investigated and analyzed, using magnetic energy and helicity. %proxies. 
While magnetic energy, in particular free magnetic energy, exhibited strong differences during the evolution of various configurations, some helicity-based quantities were robustly discriminating between eruptive and non-eruptive simulations. The total relative magnetic helicity $H$ was not one of them, showing its largest budget for the non-eruptive simulation. 
However, the ratio $H_j/H$ showed high values for eruptive simulations only, most importantly \emph{before} the onset of eruptions. 
Shortly after eruptions, its values were similar for both eruptive and non-eruptive simulations. As a result, the ratio $H_j/H$ was portrayed as a promising eruptivity proxy. 

Given the above, the actual predictive value of the ratio $H_j/H$ will ultimately be determined by data-driven modeling of observed solar ARs. 
If future computational advances make the meaningful calculation of the non-potential 3D coronal magnetic field feasible in NRT, testing of this proxy will become possible and may lead to innovative flare predictors in future, advanced versions of FLARECAST and other operationally-oriented systems. 

\begin{figure}
    \centering
    \includegraphics[width=\textwidth]{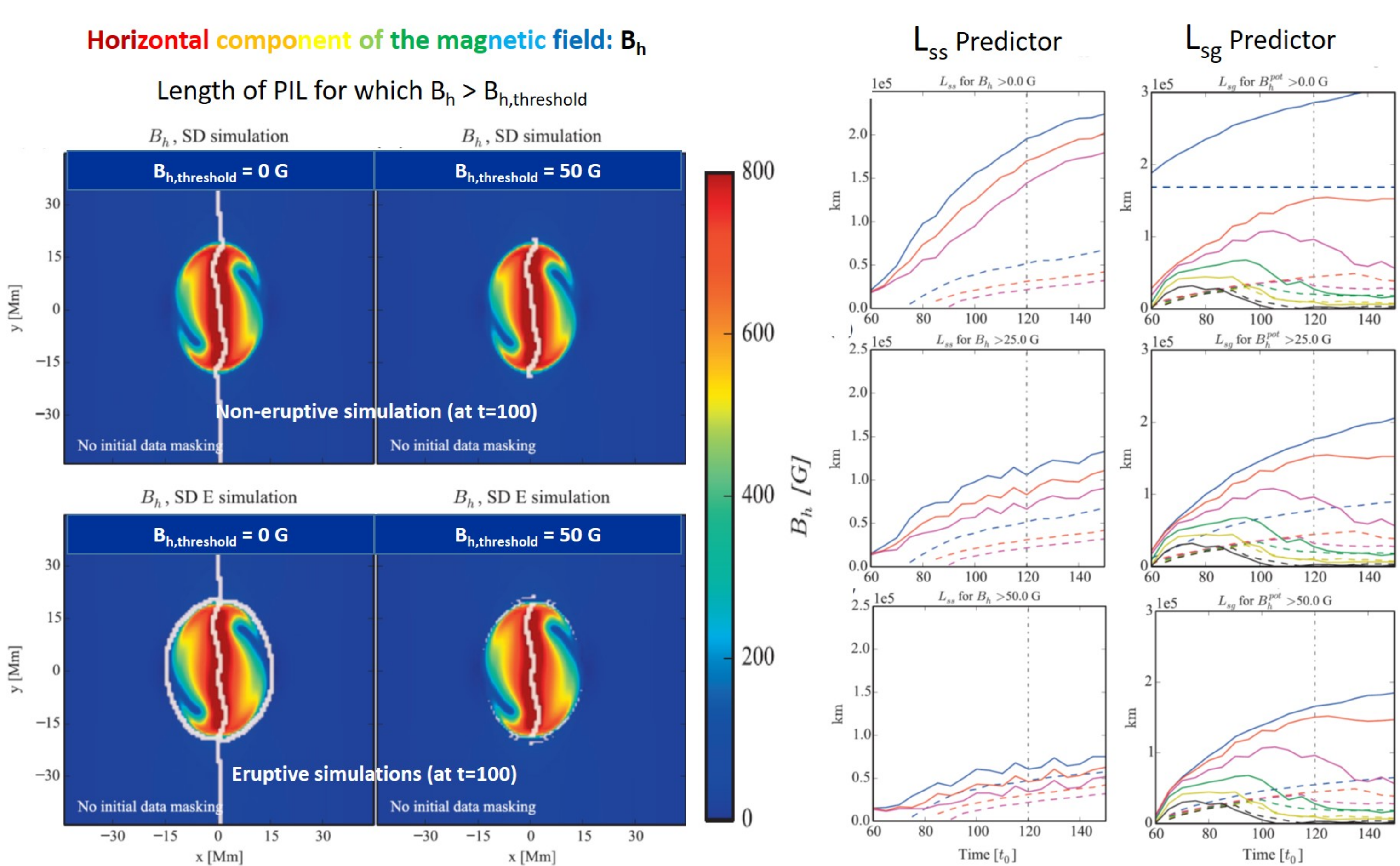}
    \caption{Example of the impact of thresholding, here of the threshold $B_{h,th}$ for the horizontal magnetic field strength $B_h$, on the efficiency of two potential flare predictors, namely the lengths of the strong-shear and strong-gradient PIL, $L_{ss}$ and $L_{sg}$, respectively (see text for details). Figure adapted from \citet{Guennou17}.}
    \label{fig:Guennou17}
\end{figure}

Synthetic data obtained from numerical simulations can also be used to study flare proxies following the same methodology with observed data. 
\citet{Guennou17} studied flux emergence simulations by  \citet{Leake13,Leake14} in a way that was similar to the treatment of observational data, namely by isolating the 2D magnetic field vector of the modeled AR photosphere. 
A list of about 100 different scalar properties was then computed in a way similar to the procedures developed in FLARECAST. 
\citet{Guennou17} found that, among the properties tested, 
only those associated with PIL features presented significant preflare signatures.
%the only ones that presented significant preflare signatures were those quantifying the total non-potentiality of ARs associated with PIL properties. 
Virtually all other properties were not able to differentiate between eruptive and quiescent simulations. 
Since the computation of several of these properties depends on the choice of parameters (e.g., thresholds for masking the data, magnetic field thresholds, etc.), \citet{Guennou17} could study the role of these thresholds on the ability of a property to determine the eruptivity. As exemplified in Figure~\ref{fig:Guennou17}, it was found that some properties were highly sensitive to the choice of user-defined thresholds whereas some others were more robust. The four panels on the left show the distribution of $B_h$ at the modeled photospheric boundary of two 3D MHD flux emergence simulations at t=100, in normalized time units of the non-eruptive (top) and eruptive (bottom) simulations. The white curves show the portion of the PIL where $B_h>B_{h,th}$, for two different choices of $B_{h,th}$ (0~G and 50~G). The length of the PIL where $B_h>B_{h,th}$ is one of two criteria used in the computation of $L_{ss}$, while  $L_{sg}$ uses a different criterion based on the potential field, namely the length of the PIL where $B_{h,pot}>B_{h,th}$.

%The left panels aims to show a criterion with a possible large impact on the computation of $L_{ss}$.  
The six plots on the right show the temporal evolution of $L_{ss}$ and $L_{sg}$ for the eruptive (solid curves) and non-eruptive (dashed curves) simulations under three different values of $B_{h,th}$, namely 0~G (top), 25~G (middle), and 50~G (bottom). The different colors corresponds to different choices of criteria also entering in the computation of $L_{ss}$ and $L_{sg}$, with details discussed in \citet{Guennou17}.  As $B_{h,th}$ increases, $L_{ss}$ becomes similar for the non-eruptive and the eruptive simulation.  This predictor is thus highly sensitive to the choice of $B_{h,th}$. 
Conversely, $L_{sg}$ is relatively insensitive to the choice of $B_{h,th}$. As a result, $L_{sg}$ is capable of discriminating the two simulations and could conceivably be used as an eruptivity proxy.

In addition, \citet{Guennou17} studied the robustness of certain parameters in the presence of noise, reporting that parameters related to electric currents seem to be the most sensitive. This would imply the need for screening such predictors before inserting them in prediction pipelines.
%This result is also important for projects such as FLARECAST because it can help in the screening of predictive properties before they are inserted in the feature processing or prediction pipelines.

\subsection{Flare-CME connection studies} 
\label{sec:FlareCME}
%\textcolor{magenta}{Sophie: Here's a first go at the HELCATS-FLARECAST work. I could certainly go into more detail if needed and add figure (although will have to check re Solar Physics journal rules).}

%EP The Explorative Research Work Package (WP6) enabled FLARECAST researchers to expand upon the results of the project into other cutting-edge research areas, including coronal mass ejection (CME) prediction. Space weather operations are currently lacking sufficient warning for solar eruptions, with forecasters only predicting CME arrival at Earth once the eruption is observed. CMEs and other solar eruptive phenomena can be physically linked by combining data from a multitude of ground-based and space-based instruments as well as models, however this can be challenging for automated operational systems. Understanding the processes and sources involved in the solar eruptions that cause extreme space weather at Earth events is imperative to improve this area of forecasting. 

The FLARECAST exploratory component further enabled researchers to expand upon the project results into other cutting-edge research areas, including coronal mass ejection (CME) prediction. 
Space-weather operations are currently lacking sufficient warning for eruptive flares, with forecasters only predicting CME arrival at Earth once an eruption is observed \citep[see, e.g.,][]{mostl_etal14,verbeke_etal19}. 
One may reasonably expect that CMEs and other solar eruptive phenomena can be physically linked by combining data from a multitude of ground- and space-based instruments, as well as models. 
However, this can be challenging for automated operational systems. 
Understanding the processes and sources involved in solar eruptions that could enable the use of photospheric properties is imperative for improving this area of forecasting. 

\begin{figure}
    \centering
    \includegraphics[width=\textwidth]{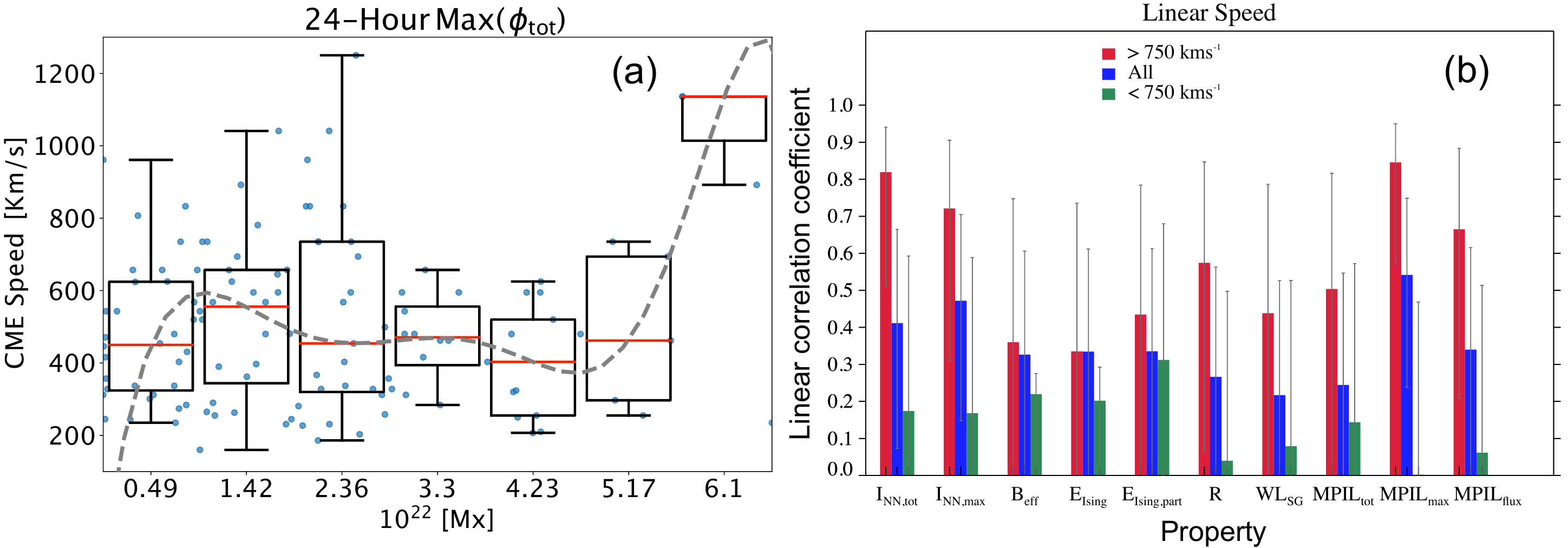} %Overleaf doesn't seem to like large file uploads so I've added a compressed pdf. You can download the original eps at this link: https://www.dropbox.com/s/kv427yb0ws38w3g/FigMurray18.eps?dl=0
    \caption{(a) Correlation between CME speed and the peak of total, unsigned magnetic flux in the source active region up to 24 hours prior to the eruption. Points correspond to the actual data values for 124 CME -- source region pairs, while a box-and-whiskers plot is overlaid along with a spline fit to quantify the correlation. Reproduced from \citet{murray18}. (b) Linear correlation coefficient and respective uncertainties between source region properties and CME speed for a sample of 32 CME -- source region pairs. A distinction between fast (red columns) and slow (green columns) CMEs has been made while results for the entire CME sample are shown in blue columns. Reproduced from \citet{kontogiannis19}.}
%    CME speed correlation with the maximum value of total magnetic flux ($\phi_{\rm tot}$) observed during the 24-hours period before an eruption. A box-and-whiskers plot with a fitted spline function overlaid show that a positive, non-linear correlation between the variables (reproduced from \citet{murray18}) Right: Figure from \citet{kontogiannis19}} }
    \label{fig:murray18}
\end{figure}

The flare-CME connection that could advance CME prediction was investigated via a unique synergy between FLARECAST and an EU Framework Programme 7 project funded under the \emph{`Exploitation of space science and exploration data'} Call. 
The HELCATS project created numerous data products from heliospheric imaging onboard the two NASA STEREO spacecraft in order to track the evolution of CMEs in the inner heliosphere. 
Using the project’s main catalogue of over 2,000 CME events imaged between 2007 and 2017, an automated algorithm was developed to connect the CMEs observed by STEREO to any corresponding solar flares and AR sources on the solar surface, resulting in the HELCATS LOWCAT catalogue. 
\citet{murray18} supplemented the LOWCAT catalogue's information on CME kinematic properties (such as speed and angular width) with FLARECAST's %extensive database of complex magnetic field 
extensive property database derived from vector magnetograms, enabling a deeper study into the characteristics of eruptive ARs. 
Initial statistical analysis was undertaken on the new combined data set, with total unsigned flux, vertical current density and current helicity identified as properties of interest for potential eruption warning thresholds. 
%The automated method developed to create the LOWCAT catalogue may also prove useful for future efforts to build an operational CME predictive capability. 
Beyond the scientific gain, this collective effort provided an excellent opportunity to foster communication between different, large-scale research projects which may work in parallel on similar research topics to benefit space-weather forecasting. 
See Figure~\ref{fig:murray18}a for an example of this synergy, while further details can be found in \citet{murray18}. 

%\textcolor{magenta}{Ioannis: here's a first draft summarizing our contribution to WP6 on CMEs.}
A further detailed investigation was performed for a smaller sample of 32 eruptive flares (4 C-class; 16 M-class; 12 X-class) by \citet{kontogiannis19}, seeking correlations between selected AR non-potentiality properties and CME kinematic characteristics. 
The selection intentionally focused on PIL-related properties, including some well-established ones (such as Schrijver’s $R$, $WL_{SG}$ and $B_{\mathrm{eff}}$) along with ones implemented in the course of FLARECAST (such as the Ising energy and the total non-neutralized electric current; see Table~\ref{tab:FC_props}). 
CME information was collected from NASA's DONKI database, utilizing CME images by LASCO onboard the SOHO mission. 
Positive correlations between CME kinematic properties were found for all pre-eruption photospheric properties, in line with the results of \citet{murray18}. 
Remarkably, correlations were stronger for faster CMEs implying that selected non-potentiality properties may be indicative of upper-limit eruption scales that ARs can produce over a given timescale. 
Among the FLARECAST properties tested, the preflare total unsigned non-neutralized current and the length of the main PIL stand out, exhibiting significantly stronger correlations with the CME 
kinetic energy and speed (Figure~\ref{fig:murray18}b). 
This finding points to a causal relationship between non-neutralized electric currents (stemming from compact, sheared PILs) and the associated free magnetic energy and helicity budgets in ARs and CMEs. 
Therefore, one envisions a future forecasting capability of basic CME kinematic properties \emph{before} CMEs actually occur that, combined with the potential CME launch location as per its source AR, could give educated guesses of possible CME shock formation and geoeffectiveness.  

\section{Conclusions}
\label{s6}

FLARECAST was a multifaceted, interdisciplinary project. 
As such, it is appropriate to identify the different activity areas of the project and summarize conclusions pertinent to each of them. 

\subsection{Science and technology}
One of the project's main aims was to seek the most efficient ways of forecasting solar flares. 
In this respect, FLARECAST is arguably the most intensive and systematic solar flare forecasting effort to date due to (i) the amount, cadence and validation of the input SDO/HMI SHARP magnetogram data, (ii) the sheer number of AR properties treated as flare predictors \citep[209 of them, with 171 used in the comprehensive study of][]{2019ApJ...883..150C}, (iii) the breadth of machine-learning prediction algorithms implemented (14 included in Table~\ref{tab:ML-methods}), and (iv) the comprehensive performance verification approach and infrastructure, along with suggestions on how to implement it in a rigorous training and testing philosophy. 

Solar flares are long known to be stochastic events, namely responses of strongly nonlinear (quite likely self-organized critical) dynamical systems, as solar ARs are thought to be. 
Hence, flare forecasting appears intrinsically probabilistic, notwithstanding that some machine-learning prediction methods provide a binary output. 
Despite this long-standing knowledge, the FLARECAST Consortium was keen to determine whether a combined Big Data and machine learning effort could lift this stochasticity barrier by achieving a compelling, verified binary separation between flaring and non-flaring AR populations. 
As \citet{2019ApJ...883..150C} declared right from that paper's title, and was also implied by the results and discussion of \citet{Florios:18}, stochasticity in flare occurrence and subsequent forecasting remains uncompromising. This, of course, could be due to limitations or shortcomings of the tested methods and / or data. Regardless, despite significant progress brought by FLARECAST and many other efforts worldwide, a decisive (i.e., binary and with high enough verification metric values to be used for decision making on an operational context) flare prediction capability is still beyond reach. This said, FLARECAST has generated codes, data and infrastructure that could potentially facilitate this crucial milestone in the future.

A recent trend in the literature \citep[see, e.g.,][and follow-up works]{nishizuka_etal18,li_etal20} suggests that machine learning may not even be sufficient, so the community should be moving toward deep learning flare prediction efforts. 
This remains to be seen; even in the above deep learning efforts, however, a significant degree of stochasticity persists, reflected in less-than-optimal performance indicators. 
So far, solar deep-learning prediction methods have by no means outperformed machine learning ones. 
It is an open question whether this can be reversed in the future, although the massive data needs of deep learning algorithms \citep[e.g.,][]{goodfellow_book}
may render flare prediction a less-than-optimal domain for deep learning efforts due to the lack of sufficient event instances to train on.

In terms of performance verification, while FLARECAST created a detailed apparatus (Section~\ref{s4}), the nominal duration of the project did not allow its complete exploitation. 
Existing project studies \citep{benvenuto2018hybrid, Florios:18,2019ApJ...883..150C} have used meaningful parts of it but not the entire infrastructure provided in %Figure~\ref{fig:JSON} and 
Table~\ref{tab:bmetrics}. 
Future studies, however, will have the option to work and rely on this infrastructure for a comprehensive performance verification assessment.

While FLARECAST was a case study of R2O, it soon became clear that we could learn from its output, by means of both predictor ranking and the explorative research component. 
Then, part of the project's output contributes to O2R objectives. 
In  particular, a robust ranking of predictors triggers the question of why just a few ($\lesssim$ 10) predictors,  
different for each prediction method at least for flares of GOES class $\geqslant$M1.0, gather virtually all the predictive power
in \citet{2019ApJ...883..150C}, while in 
another attempt by \citet{Florios:18} that used the same (RF) method the increase in performance seems monotonic as more predictors are added. The above  generalize and diversify the big picture that now calls for (i) a comprehensive review of training and testing practices, regardless of prediction methods, and (ii) an investigation of specific predictors' performance on given prediction method(s). In light of all these, one
%\citep{Florios:18,2019ApJ...883..150C}. 
%The project solidified the knowledge that different machine learning methods often give rise to different \mg{predictor} ranking \citep{Florios:18,2019ApJ...883..150C}. 
%This generalizes the picture that now calls for investigating performance of certain predictors as per given prediction method(s), a dependency that should be studied further in the future. 
key takeaway message is that we should not be employing `black box' (i.e.,  non-human-interpretable) prediction methods, in spite of their performance, in order to keep the knowledge discovery channel open \citep{rudin19}. 

In terms of exploration, the project used synthetic data \citep{Guennou17,Pariat17} to investigate elements not accessible by the available observational data, but also concerned itself with the flare-CME connection to investigate flare predictors exhibiting a statistically significant correlation with CME parameters. This was pursued via both the FLARECAST -- HELCATS synergy \citep{murray18} and the FLARECAST -- DONKI association of \citet{kontogiannis19}.
These works, along with previous ones in this respect \citep{falconer_etal02,barnes07,uura_etal07,georgoulis08}, all show promise toward a viable interpretation of the pre-eruption configuration in solar ARs and a deeper, more complete understanding of solar eruptions. 
%This can serve another future O2R purpose, facilitated by the combined study of FLARECAST's results and those of other projects such as HELCATS. 

Along with other recent studies \citep{barnes_etal16,leka_etal19a,leka_etal19b,park_etal20}, FLARECAST has showcased the need for fixed, pre-defined data sets in solar flare (and eruption) forecasting. 
Such `benchmark' data sets \citep[see][for a definition and an example]{angryk_etal20} enable the precise evaluation of different methods, machine / deep learning or other, on preset training and testing samples. Performance comparisons on different data sets and / or phases of the solar cycle can be problematic due to the differences in underlying flare statistics over different time periods, so a benchmark dataset should be partitioned in such a way as to treat variable climatology to the extent possible. The FLARECAST property database could also play the role of a benchmark dataset. It further enables the important action of time series forecasting, that was unfortunately not achieved during the project's nominal period of execution, and the treatment of class imbalance in machine learning performance. 
%that is already under independent study \citep{ahmadzadeh_etal20}. 
Such investigations directly address another trait of major flares, namely their scarcity and rareness, that becomes extreme for flares of historic magnitudes. 

FLARECAST developed an infrastructure that 
could help address the need for an integrated space weather forecasting platform. 
Coordination of space weather forecasting efforts has been identified as a key priority in proposed roadmaps such as the one developed on behalf of the COSPAR \citep{schrijver_etal15}. 
It is also prominently recommended by the 
ESWACC \citep{oopgenoorth_etal19}. 
In line with these studies, global coordination of space-weather forecasting efforts sprung out as a top-level conclusion during the FLARECAST Stakeholder Workshops. 
The project's infrastructure could serve as a testbed, or breadboard, for future applications encompassing the other two main legs of ``stormy'' space weather, namely CMEs and SEPs. 
Given the versatility of the Docker engine and containers used in FLARECAST, components of different programming languages can be removed, updated, or added in a highly modular fashion. 

All in all, FLARECAST has amply shown that solar flare forecasting should not be, and thankfully is not, an internal affair for heliophysicists. 
Fusion of expertise is paramount to advance and, ultimately, break ground toward an efficient space-weather forecasting but also other complex real-world problems. 

\subsection{Operations}
By defining different flare forecasting modes (i.e., based on GOES flare class, forecast window, latency, etc.) and testing numerous prediction algorithms, FLARECAST reaffirmed that one single recipe does not fit every need. 
Patterns between different prediction methods were seen in the best-performing predictors for GOES $\geqslant$C1.0 flares (i.e., flares of GOES C-class and above), for example, but not for GOES $\geqslant$M1.0 flares  \citep{2019ApJ...883..150C}. 
In addition, AR properties that correlate best with CME properties \citep{murray18, kontogiannis19} may, or may not, be within the best performing predictors for a given method. 
We view these findings in light of an outcome of the FLARECAST Stakeholders Workshops that the real end users of the project's results are most likely expert operational forecasters who can disseminate them as needed. 
Other synergistic community works fall along similar lines by arguing for the benefit of human `forecasters in the loop' \citep{leka_etal19b}. 
It then becomes clear that more time and effort will be needed before agencies and institutions committed to operations are able to shift from today's relatively simple forecast philosophies to sophisticated, multi-dimensional parameter spaces, Big Data and intensive machine / deep learning methodologies. 
It is a crucial step to be taken, however, balancing on a fine line between simplicity and interpretability of the concepts and the necessary level of complexity dictated by the problem at hand. 

In brief, in spite of its limited duration that left a few envisioned elements to be realised (i.e., time series forecasting and comprehensive performance verification of all prediction algorithms), we hope that FLARECAST can be thought of as a future `textbook' R2O case, paving new ways and even narrowing `the valley of death' of less-than-successful approaches between the two domains of research and operations. 
As eloquently put by \citet{merceret_etal13}, who drew analogues between space weather and terrestrial meteorology, there may be leads able to transform the `valley of death' into a `valley of opportunity'.

It should not go unnoticed that the immediate next step for the FLARECAST infrastructure is to become a flare prediction service of ESA's SSA Space Weather Service Network\footnote{Available at \url{https://swe.ssa.esa.int/ssa-space-weather-activities}}. This task is undertaken by the FHNW partner that has originally developed the project's technology apparatus.

\subsection{Communication}
A key objective of the project was the communication with government and industry representatives, that cultivated into two FLARECAST Stakeholder Workshops. 
A top-level conclusion of them was that communities involved in (solar flare or space-weather) forecasting should be in close contact. 
There are `language' and terminology barriers that require a continual effort toward keeping the momentum of communication and better defining the constantly evolving landscape of user needs. 
A prominent example in this direction is the aviation and power industries that have moved beyond colloquial interest and are raising their awareness toward the actual capabilities, shortcomings and detailed products of operational forecasting. 

We learned from these Stakeholder Workshops that accuracy, timeliness, and ease of use are the most important factors in a forecast from the users' perspective. Less important is the scientific detail accompanying each forecast. 
As mentioned above, it was also found that there should be an intermediate step between the researchers who devise the forecasts and the end users. 
This step is filled by operational forecasters, who can interpret forecasts and convey their messages in a language that users can understand and cope with. 

As to the communication with the wider public, FLARECAST reaffirmed that the public is keen to listen and eagerly seeks knowledge on solar flares and space weather. 
It became clear that open-access, public-domain projects such as this one should return part of their discovered knowledge to the taxpayer citizen, by educating them credibly and responsibly, steering clear from the (unfortunately high) level of misinformation spread largely untested and un-scrutinized over the internet and social media\footnote{As a side note, recent efforts to counter the spread of fake news in the internet also rely heavily on the use of (interpretable) deep learning architectures \citep[e.g.,][]{kai_etal19}.}.

\subsection{Lessons learned}
In implementing a sizable and diverse project such as FLARECAST, there was no shortage of lessons to learn. 
The intrinsic complexity of the project was not only due to its diversity, but also because Consortium partners were geographically distributed throughout Europe. 
This required certain steps to be taken in order to secure efficient coordination and implementation. 
Even so, however, some desired tasks could not be implemented within the nominal duration due to lack of time. 
The most important lessons were the following: 
\begin{itemize}
    \item  An effective project governance is instrumental for planning ahead. 
    Well-defined, distinct roles foreseen in the Consortium Agreement such as those of the Project Coordinator, Project Scientist, Financial and Legal Manager, as well as bodies such as the Project Management Board and the Steering Committee, provide a governance structure that can, in principle, lead the project into fruition via a properly shared workload. 
    There should be space for some  adjustments in the initial plan. However, radical changes midway through the project will only lead to confusion and impede implementation.
    \item The Consortium further needed an efficient way of remote communication, beyond e-mails, phone calls and teleconferences. 
    This was achieved via an integrated platform of collaborative software tools that were utilized on a Consortium-wide level. 
    This system combined both flexible internal wiki-pages (e.g., enabling tagging individuals in documents with associated email updates, real-time recording of meeting minutes, efficient exporting of periodic report documents and code documentation, etc.) with integrated code repositories that have easily interpretable visualization interfaces for version tracking (hence, allowing efficient code reviews and prompt troubleshooting of change-related issues).  
    \item In Consortium deliberations, common understanding and a continual pursuit of consensus in a collegial atmosphere were valued as the most important elements for a successful collaboration.
    \item In a project with strict requirements for hardware installation and software development, time (and backup time) is necessary; efficient time management is paramount for the aversion of implementation delays. 
    Given the high degree of synchronization required between parallel tasks, delays often seem inevitable; when they happen, they may also cause a domino effect affecting different project parts. 
    Advance risk assessment and time planning help mitigate the impact of these delays.
\end{itemize}
\begin{acknowledgements}
The FLARECAST project was funded by the European Union Research and Innovation Programme under Grant Agreement no. 640216. 
The project would not have been possible without NASA's SDO mission and the unrestricted access to SDO data, primarily from Stanford University's JSOC. While we wish to sincerely thank all community members who participated in the public discussions, debates and the Stakeholder Workshops, the FLARECAST Consortium is particularly grateful to the EU-assigned Project Reviewer, Thierry Dudok de Wit, and the EU Project Officer, Sabri Mekaoui, for their invaluable and continuous help and support. The project's administrative component has benefited enormously by the administrative structures of all partners, coordinated flawlessly by the Project Manager, Vangelis Argoudelis. 
We are also indebted to the FLARECAST Steering Committee (Neal Hurlburt [Chair], Graham Barnes, Douglas Biesecker, Pedro Russo, and Silvia Villa) who devoted time and effort on the project and contributed decisively to its smooth implementation. Finally, we deeply appreciate the contribution of the two anonymous reviewers whose insightful comments helped us substantially improve the presentation, content and context of the original manuscript.
\end{acknowledgements}

\bibliography{references_1,refs_mkg}

\begin{appendix}

\section{Open access of FLARECAST data, codes and infrastructure}
\label{app1}

\subsection{Data}
Access to the FLARECAST property database is provided via \url{https://api.flarecast.eu/property/ui/}, while the prediction database is accessible via \url{https://api.flarecast.eu/prediction/ui/}. 
Both services are Swagger-based APIs, enabling data access in both human and machine readable format. 
They provide a collection of so-called API routes to retrieve and analyse the FLARECAST data. 
While analysis is useful for developers and system administrators, the \emph{View} routes are intended for end users. 
The graphical interfaces at these URLs serve two purposes. 
First, one can retrieve and verify data manually. 
Second, the interfaces print out a URL than can be used as query template and can also be integrated in any programming language supporting HTTP file access. 

The most useful \emph{View} routes of the property database service are: 
\begin{itemize}
    \item \emph{/data set/list}, that provides the list of all registered data sets. 
    FLARECAST primarily works with the \emph{production\_03} (1 September 2012 to 12 April 2016 and 9 September 2017 to 30 January 2019) and \emph{questionable\_02} (13 April 2016 to 8 September 2017) data sets. 
    The second data set includes HMI NRT data that may be problematic, according to a JSOC release\footnote{{\url{http://jsoc.stanford.edu/jsocwiki/ModLRecalibration}}}. 
    \item \emph{/property\_types}, that provides the list of individual property names. 
    This is useful when one is interested in the acquisition of certain properties only. 
    \item \emph{/data set/\{data set\}/list}, that provides the main access to the property database. 
    Various options can be used to refine the search, while blank data requests are also possible. 
    As an example, a general API URL for all property values within the 2-day interval of 1 June 2014 to 23 June 2014 at all available cadences reads  
    
    \url{https://api.flarecast.eu/property/region/production_03/list?cadence=all&exclude_higher_cadences=false&time_start=between(2014-06-01,2014-06-03)&property_type=*&region_fields=*}.
    
    One may refine the search as needed, in terms of \emph{cadence}, time interval (via \emph{time\_start}), or specific property (via \emph{property\_type}). 
    %MSO: I would just drop this entry. \item The route \emph{/data set/\{data set\}/summary} provides a graphical summary of generated properties and cadences. This typically takes significant time to be build so it is generally not recommended. 
\end{itemize}
The most useful \emph{View} route of the prediction database service is: 
\begin{itemize}
    \item[$\bullet$] \emph{/prediction/list\_v2}, that provides the main access to the prediction database. As with the property database, various options can be used to refine the search. As an example, an API URL for all available predictions within the 5-month interval of 1 January 2018 to 1 June 2018 reads
    
    \url{https://api.flarecast.eu/prediction/prediction/list_v2?include_flare_associations=true&algorithm_config_version=latest&prediction_time_start=between(2018-01-01,2018-06-01)}.
    
    The search window can be adapted by \emph{prediction\_time\_start}. 
    Each entry provides the algorithm configuration name (\emph{algorithm\_config\_name}) and id (\emph{algorithm\_config\_id}), along with all other applicable information. 
    One can look into certain algorithm configuration names to refine the search further by setting the \emph{algorithm\_config\_name} field accordingly and obtain the respective API route. 
\end{itemize}
Most other routes in the property and prediction database services are still used by some of the user interfaces and some debugging scripts, but are not intended for end users. 

\subsection{Codes} 
All FLARECAST source codes and computational infrastructure are public and can be found in the project's Bitbucket repository: \url{https://dev.flarecast.eu/stash/projects/}. 
The repository includes different directories, namely: 
\begin{itemize}
    \item {\bf Download:} algorithms for downloading SWPC and flare staging data. 
    \item {\bf Property Extraction:} codes and algorithms for the extraction of all properties included in Table~\ref{tab:FC_props}. 
    \item {\bf Flare Prediction:} source codes for the machine learning algorithms included in Table~\ref{tab:ML-methods}.
    \item {\bf FLARECAST Data Model:} codes for defining and implementing the project's data model (Section~\ref{sec:dm}). 
    \item {\bf FLARECAST Infrastructure:} codes for the project's computational infrastructure (Section~\ref{s3}). 
\end{itemize}
Algorithms typically include brief usage notes, examples and information on each applicable programming language. 

\subsection{Infrastructure installation}
The first dependency is the installation of a Docker engine from \url{https://www.docker.com}. 
Docker is an open source software that offers an ecosystem in which different Docker containers co-exist and function independently. 
Given that each FLARECAST Docker container contains a different routine or algorithm written in different programming languages (IDL, Python, C, or R), the second dependency is that the installation server must be equipped with licenses of these languages to be able to edit / modify the infrastructure. 
%DSB: Has it been checked that this sequence of scripts still works?
With Docker engine installation in place, the FLARECAST infrastructure is installed in a Unix/Linux command line environment as follows: 
\begin{itemize}
    \item Download the installer script:\\ 
    \url{curl} \url{-o} \url{infrastructure.sh}  \url{https://dev.flarecast.eu/stash/projects/INFRA/repos/dev-infra/browse/infrastructure.sh?raw} 
    \item Assign execution mode to the script:\\
    \url{cmod} \url{a+x} \url{infrastructure.sh} 
    \item Setup the infrastructure or update to the most recent version:\\
    \url{./infrastructure.sh} \url{update} 
    \item Finally, run the development infrastructure:\\
    \url{./infrastructure.sh} \url{run} 
\end{itemize}

\subsection {Licensing}
\label{sec:license}
EU's OpenAIRE policy requires that all source code and data models must be openly accessible to any interested person or entity, both within the EU and worldwide. 
Therefore, the software and data licenses chosen by the FLARECAST Consortium aim to be as permissive as possible, at the same time being free of any liability and provided `as is'. The majority of the written code is available under a BSD-2 (\url{https://opensource.org/licenses/BSD-2-Clause}) license. Some of the algorithms, however, depend on third-party libraries that use a less permissive GPL licence (\url{https://www.gnu.org/licenses/gpl-3.0.de.html}) and thus these algorithms need to be released under GPL, as well. 
All FLARECAST data products are provided under the fully open ODC PPDL (\url{https://www.opendatacommons.org/licenses/pddl/1-0/index.html}) license. 

\section{The FLARECAST User Survey}
\label{app2}
The FLARECAST user survey was carried out in Autumn 2016, with around 100 users approached and 31 responding. 
Around 60\% indicated that they were using flare forecast or alert services. 
Most thought these services were useful but around a quarter did not know whether their preferred service was accurate or inaccurate, which points to a need for better education. 
Asked whether they would recommend such a service, around two thirds gave a ``passive'' response, indicating that they neither strongly recommend, nor strongly not recommend, flare forecasting services.

Users were asked if they planned to use flare forecasting services differently in the future. 
Five of them provided further details of how they would do this. 
While the assumption is that most users are in the aviation or defence industries, it is interesting that out of these five comments, one was from a satellite user and two from GNSS users. 
Suggested approaches included:
\begin{itemize}
    \item Developing an operational response to space weather events, and a change in practices to take advantage of increased precision and advance warnings.
    \item Comparison with other flare prediction tools, such as ASSA{\footnote{\url{http://spaceweather.rra.go.kr/models/assa}}}.
    \item Examining correlations between flares and ionospheric TEC changes.
    \item Checking the impact of space weather disturbances on GNSS systems.
    \item Allowing the spacecraft operator to tailor services for spacecraft control aspects.
\end{itemize}

\begin{figure}
    \centering
    \includegraphics[width=\textwidth]{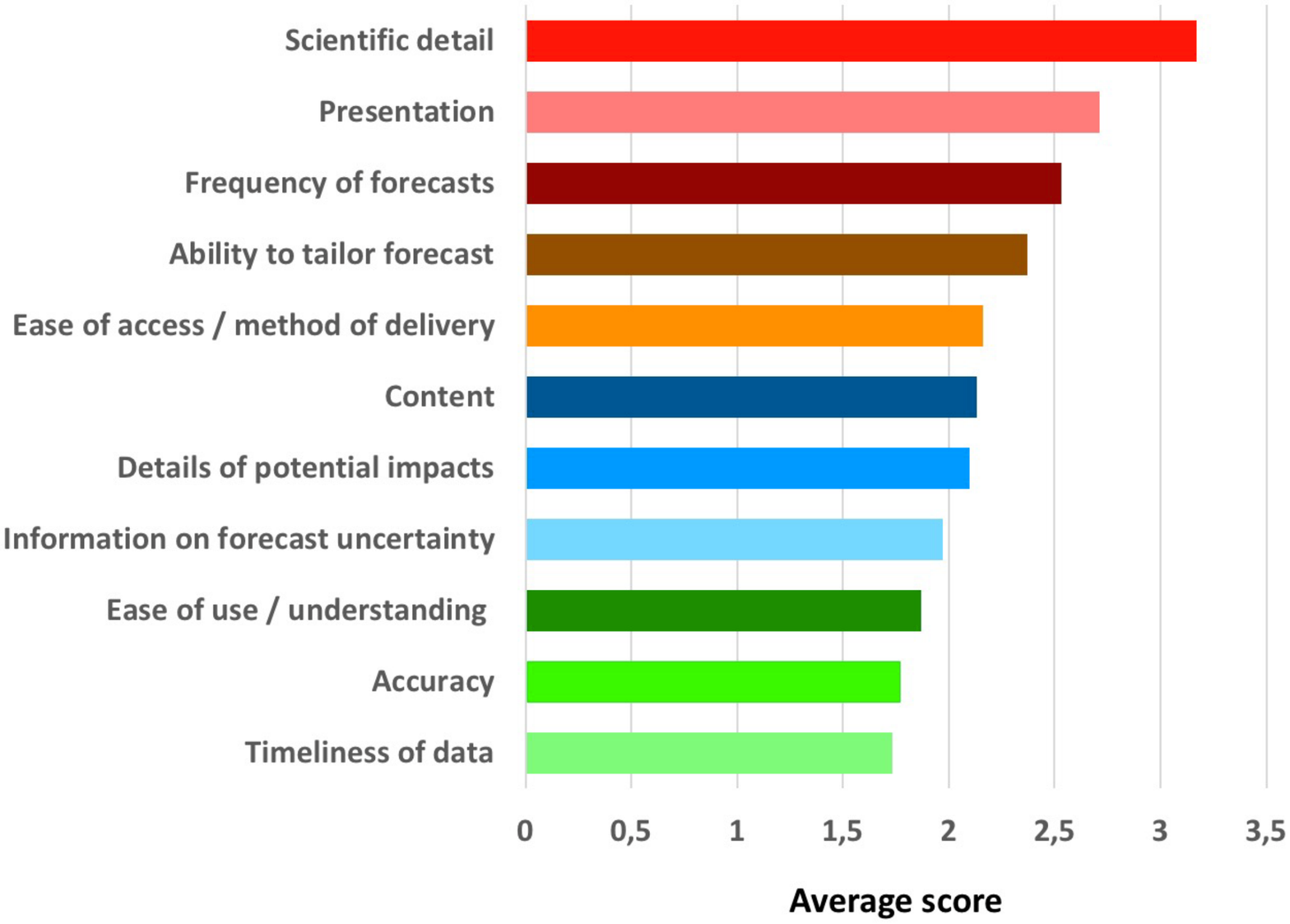}
    \caption{Summary of responses to the question, ``Which factors are/would be important to you in a flare forecasting service?'' 
    Lowest average score corresponds to the most important factor.}
    \label{fig:SurveyFig}
\end{figure}
The users were then asked what factors would be most important to them in a flare forecasting service. 
Timeliness and accuracy were found to be most important to a user, while scientific details were deemed least important. 
A summary of all responses appears in Figure~\ref{fig:SurveyFig}.

Most flare forecasts focus on the occurrence of flares defined by the NOAA R scales (e.g., M-class, X-class), but 77\% of responders said forecasts of ``All Clear'' periods are also important. 
A similar number knew about the NOAA R scales, but 12.5\% said that they were not useful for their purposes. 
The following reasons were given:
\begin{itemize}
    \item[$\bullet$] The R scales classify impacts to radio systems, so are of some interest to us (a Telecoms company), but we are also interested in space weather impacts to other systems such as power distribution.
    \item[$\bullet$] The scale is useful globally, but too coarse at local (i.e., country) level.
    \item[$\bullet$] Currently we are not reacting to NOAA scales. A decision was taken to react to MOSWOC sourced indices.
\end{itemize}

Finally, the survey came up with suggestions of other useful points for discussion during the First Stakeholders Workshop. Suggestions included:
\begin{itemize}
\item	Details on verification of results and methods. 
\item	Explanations for non-scientists; international coordination to raise awareness.
\item Coupling to D-RAP{\footnote{\url{https://www.swpc.noaa.gov/products/d-region-absorption-predictions-d-rap}}}; specific solar radio burst forecasts/alerts. 
\item	The state of the science and roadblocks for forecast developments.
\item	Better understanding of impact / consequence of space weather events, and presentation of this for non-experts.
\item	Clear presentation of applicable uncertainties.
\end{itemize}

\section{FLARECAST-related refereed publications}
\label{app3}
A total of eighteen (18) refereed publications, including this overview, were fully or partially supported by the FLARECAST project. The other seventeen (17) are (alphabetically by first author): 
\citet{barnes_etal16, benvenuto2018hybrid,2019ApJ...883..150C,Florios:18,Gontikakis_etal20,Guennou17,Guerra2018,kontogiannis17,kontogiannis18,kontogiannis19,massone_etal18,mccloskey_etal16,murray17,murray18,Pariat17,park18,sharpe_murray17}.

\newpage

\section{List of acronyms}
\label{app4}
\begin{tabular}{l l}
2D & : Two-dimensional\\
3D & : Three-dimensional\\
ACC & : Accuracy\\
AD & : Appleman's discriminant\\
AIA & : Atmospheric Imaging Assembly\\
API & : Application programming interface\\
AR & : Active region\\
ASSA & : Automatic Solar Synoptic Analyzer\\
AUC & : Area under curve\\
BIAS & : Bias\\
BS & : Brier score\\
BSS & : Brier skill score\\
CCMC & : Community Coordinated Modeling Center\\
CME & : Coronal mass ejection\\
CNRS & : Centre national de la recherche scientifique\\
COSPAR & : Committee on Space Research\\
D-RAP & : D region absorption prediction\\
DB & : Data base\\
DEM & : Differential emission measure\\
DNA & : Deoxyribonucleic acid\\
DONKI & : Space weather database of notifications, knowledge, information\\
EASA & : European Aviation Safety Agency\\
EM & : Emission measure\\
ESA & : European Space Agency\\
ESWACC & : European Space Weather Assessment and Consolidation Committee\\
ETS & : Equitable threat score\\
EU & : European Union\\
EUV & : Extreme ultraviolet\\
FAR & : False alarm ratio\\
FHNW & : Fachhochschule Nordwestschweiz\\
FITS & : Flexible image transport system\\
FLARECAST & : Flare likelihood and region eruption forecasting\\
FN & : False negative\\
FP & : False positive\\
FOV & : Field of view\\
GAIA & : Gaussian AIA\\
GNSS & : Global Navigation Satellite System\\
GOES & : Geostationary Operational Environmental Satellites\\
GPS & : Global Positioning System\\
HARP & : HMI active region patch\\
HEK & : Heliophysics Events Knowledgebase\\
HELCATS & : Heliospheric cataloguing, Analysis and Techniques Service\\
\end{tabular}
\newpage
\begin{tabular}{l l}
HMI & : Helioseismic and Magnetic Imager\\
HSS & : Heidke skill score\\
HTTP & : Hypertext Transfer Protocol\\
IDL & : Interactive data language\\
JSOC & : Joint Science Operations Center\\
JSON & : Javascript object notation\\
LASCO & : Large Angle and Specrometric Coronagraph\\
LASSO & : Least absolute shrinkage and selection operator\\
LOS & : Line of sight\\
LOWCAT & : Low corona catalog\\
MEDOC & : Multi Experiment Data \& Operation Center\\
MHD & : Magnetohydrodynamics\\
MLP & : Multi-Layer Perceptron\\
MOSWOC & : Met Office Space Weather Operations Centre\\
NaN & : Not a number\\
NASA & : National Aeronautics and Space Administration\\
NetDRMS & : Network data record and management system\\
NOAA & : National Oceanic and Atmospheric Administration\\
NPDA & : Nonparametric discriminant analysis\\
NRT & : Near-realtime\\
O2R & : Operations to Research\\
OR & : Odds ratio\\
ORSS & : Odds ratio skill score\\
PFE & : Potential field extrapolation\\
PIL & : Polarity inversion line\\
POD & : Probability of detection\\
POFD & : Probability of false detection\\
PostgreSQL & : Postgres structured query language\\
PROTEC & : Protection of our assets in space\\
R2O & : Research to Operations\\
RF & : Random forests\\
RNN & : Recurrent neural network\\
ROC & : Receiver operating characteristic\\
SDO & : Solar Dynamics Observatory\\
SEDI & : Symmetric extremal dependence index\\
SEP & : Solar energetic particles\\
SHARP & : Space Weather HMI active region patch\\
SOC & : Self-organized criticality\\
SOHO & : Solar and Heliospheric Observatory\\
SRS & : Solar region summary\\
SSA & : Space Situational Awareness\\
STEREO & : Solar Terrestrial Relations Observatory\\
SVM & : Support vector machine\\
SWPC & : Space Weather Prediction Center\\
\end{tabular}
\newpage
\begin{tabular}{l l}
SXI & : Solar X-Ray Imager\\
TEC & : Total electron content\\
TN & : True negative\\
TP & : True positive\\
TS & : Threat score\\
TSS & : True skill statistic\\
URL & : Uniform resource locator\\
WCS & : World Coordinate System\\
WP & : Work package\\
\end{tabular}
\end{appendix}   
   
%\end{linenumbers}

\end{document}